\documentclass[preprint,12pt,review]{elsarticle}

\usepackage{amssymb}
\usepackage{amsmath}
\usepackage{color}
\usepackage{spalign}
\usepackage{svg}
\usepackage{amsthm}
\usepackage{hyperref}
\usepackage{tabularx}
\usepackage{diagbox}
\usepackage[T1]{fontenc}
\usepackage[caption = false]{subfig}
\usepackage{multirow}

\usepackage[normalem]{ulem}  

\journal{Polymer}

\begin{document}

\begin{frontmatter}

    \title{Computational study of active polar polymer melts: from active reptation to activity induced local alignment }


    \author[inst2]{J. Oller-Iscar$^*$}
	\author[inst1]{Andr\'{e}s R. Tejedor$^*$}
	\author[inst3]{Marisol Ripoll}
	\author[inst2]{Jorge Ram\'{i}rez}
    \ead{jorge.ramirez@upm.es}
        
	\affiliation[inst2]{organization={Department of Chemical Engineering, Universidad Polit\'{e}cnica de Madrid},
		addressline={Jose Gutierrez Abascal 2},
		city={Madrid},
		postcode={28006},
		state={},
		country={Spain}}
    \affiliation[inst1]{organization={Yusuf Hamied Department of Chemistry, University of Cambridge}, addressline={Lensfield Road}, city={Cambridge}, postcode={CB2 1EW}, state={England}, country={United Kingdom}}
    \affiliation[inst3]{organization={Theoretical Physics of Living Matter, Institute for Advanced Simulation, 
Forschungszentrum J\"ulich}, addressline={52425 J\"ulich}, country={Germany}. $^*$These authors contributed equally.}

	\begin{abstract}
This work investigates the effects of tangent polar activity on the conformational and dynamic properties of entangled polymer melts through Langevin molecular dynamics simulations. We examine systems composed of all self-propelled, monodisperse linear chains, so that constraint release is considered. The range of activities explored here includes values where the active reptation theory is applicable, as well as higher activities that challenge the validity of the theory. Chain conformations exhibit a moderate increase in coil size increase, which becomes more pronounced at higher activity levels. Under these conditions, a local bond alignment along the chain contour appears together with a non-homogeneous segmental stretching, and orientation and stretching of the tube. Dynamically, polar activity induces a molecular-weight-independent diffusion coefficient, a transient superdiffusive behavior, and an end-to-end relaxation time inversely proportional to the molecular weight. Finally, our results are summarized in a diagram that classifies the various regimes of behavior observed in the simulations. Overall, these findings provide valuable insights into the complex interplay between activity and entanglements, advancing our understanding of active polymer systems and their potential applications across various fields.
	\end{abstract}

	\begin{graphicalabstract}
		\includegraphics[width=\columnwidth]{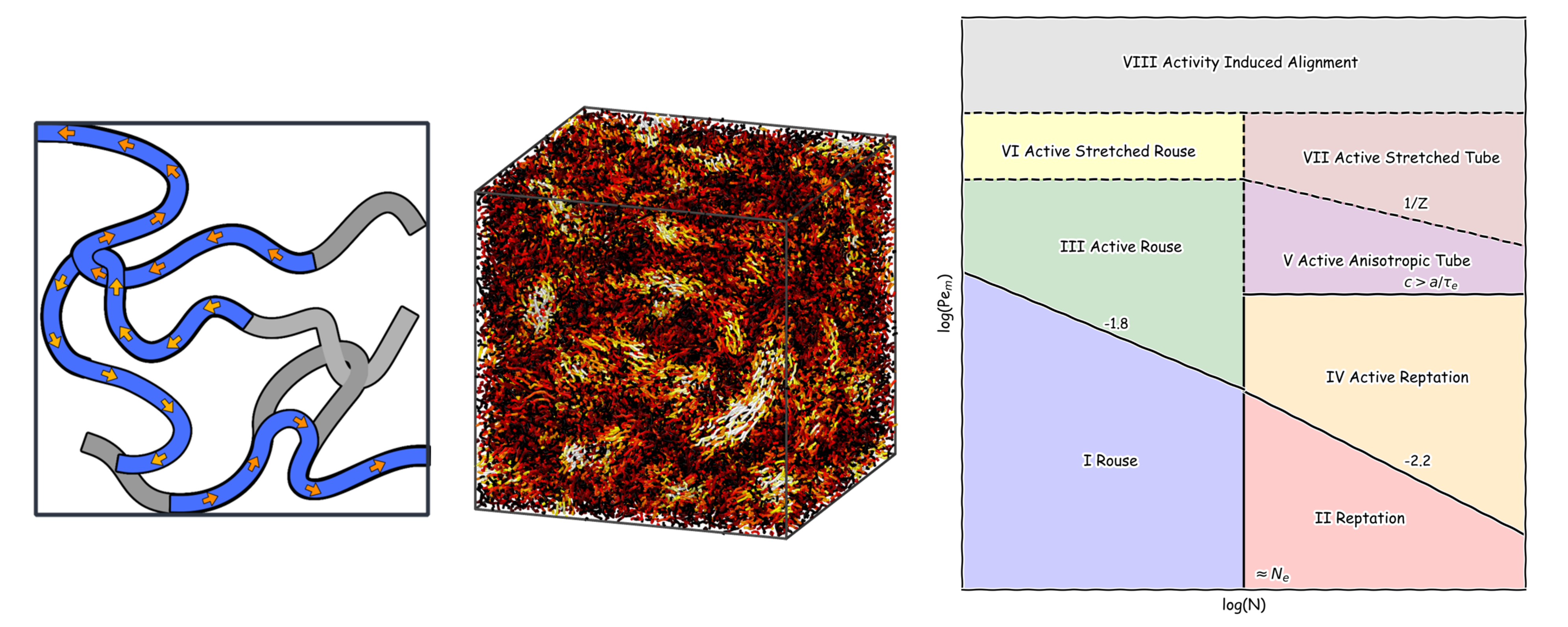}

	\end{graphicalabstract}

	\begin{highlights}
		\item Langevin molecular dynamics simulations of entangled polymers melts with polar activity 
		\item Verification of the existing theory predictions and analysis of the impact of constraint release
		\item Non-homogeneous activity-induced deformations and bond alignment along the chain contour
		\item Tube alignment and stretching at high activities
		\item Overall dynamical and conformational behavior summarized in a phase diagram
	\end{highlights}


\end{frontmatter}


\section{Introduction}
\label{sec:introduction}

Active matter comprises individual elements capable of consuming internal energy or drawing energy from their environment to perform mechanical work~\cite{Bechinger2016, gompper20202020}, leading to non-equilibrium collective behaviors not typically observed in equilibrium systems~\cite{roca2022clustering, Zottl_2016, negi2022emergent, hagan2016emergent}. There are many examples of active matter both in biological~\cite{Vicsek_2012,2004kinesin,butt2010myosin} and synthetic systems~\cite{Xu_2020, ghoshControlledPropulsionArtificial2009, jiangActiveMotionJanus2010}. Active polymers arise from the collective behavior of covalently bonded active agents~\cite{winkler2020physics}, exhibiting unique and complex dynamics that distinguish them from passive polymers. Activity in polymers can take various forms, including monomer self-propulsion~\cite{nishiguchiFlagellarDynamicsChains2018}, the action of molecular motors~\cite{dogic2020,vliegenthart2020filamentous}, or interactions with external fields~\cite{martinez-pedreroColloidalMicrowormsPropelling2015} and local gradients~\cite{jaiswalDiffusiophoreticBrownianDynamics2024}. The complex interplay between the magnitude and orientation of active forces and the internal degrees of freedom of the polymer significantly affects its conformation, relaxation time, and diffusion. Of the diverse types of active forces that can be found in polymers, polar active forces --- characterized by a tangential force that induces directed motion along the polymer chain --- have attracted significant interest for their role in biological systems like microtubules and actin molecules driven by molecular motors~\cite{ndlecSelforganizationMicrotubulesMotors1997,hirokawa2009kinesin,butt2010myosin}. These type of active forces also hold promise for synthetic biology and materials science applications due to their unique properties~\cite{schaller2010polar, philipps2022tangentially, brahmachariTemporallyCorrelatedActive2024, goychukPolymerFoldingActive2023}.

Most previous theoretical studies on active polar polymers have focused on dilute conditions~\cite{anand2020conformation,tejedorProgressivePolymerDeformation2024,martin-gomezCollectiveMotionActive2018,Bianco_2018}. Polar activity in single chains has shown to either increase or decrease overall polymer size, depending on polymer length, activity strength~\cite{Bianco_2018, anand2018structure, kumarLocalPolarLongRange2023, lamuraExcludedVolumeEffects2024, philipps2022dynamics}, hydrodynamic interactions~\cite{steijnConformationDynamicsWet2024,martin2018active}, the position of the active monomer block along the chain~\cite{vatinConformationDynamicsPartially2024}, and the relevance of inertia~\cite{tejedorProgressivePolymerDeformation2024,fazelzadeh2023effects}. In all cases, polar activity may induce a progressive deformation where the head is generally more collapsed than the tail~\cite{tejedorProgressivePolymerDeformation2024}. These findings contrast with the predictions of simulations and analytical theories for polar active Rouse chains~\cite{petersonStatisticalPropertiesTangentially2020, philipps2022tangentially}, which suggest that chain conformations remain independent of activity. This contrast underscores the importance of model details, such as excluded volume interaction and the form of bonded potentials, which can significantly impact system properties. Both simulations and theory predict a superdiffusive regime for the center-of-mass mean-square displacement, followed by an enhanced diffusive regime at longer times, similar to the behavior of active Brownian particles (ABP)~\cite{Howse_2007}. 

Previous studies have also investigated two-dimensional active polar polymer melts under dense conditions~\cite{vliegenthart2020filamentous,isele2015, duman2018collective}, revealing complex phase behaviors and non-equilibrium steady states. Ubertini \textit{et al.}~\cite{ubertiniUniversalTimeLength2024} recently examined the conformation and dynamics of polar active polymers in dense three-dimensional conditions, 
concluding that chain conformations and dynamics display universal characteristics, independent of whether chains are in dilute solutions or melts. Miranda \textit{et al.}~\cite{mirandaSelfOrganizedStatesSolutions2024} explored active flexible rings in bulk and with lateral confinement, observing the emergence of self-organized complex dynamical states. Notably, the extension to three-dimensional systems adds significant complexity --- not only computationally but also through the critical role of entanglements, which are absent in dilute conditions and less prominent in two-dimensional systems. Unraveling the interplay between activity, entanglement, and spatial dimensionality is essential for a comprehensive understanding of polar active polymers. 

In equilibrium melts, the motion of linear polymer chains with molecular weight $N$ (expressed as the number of monomers) beyond a threshold $N_c$ is constrained by the inability to cross neighboring molecules~\cite{edwards1967}. These topological constraints, or entanglements, profoundly alter the dynamics of equilibrium polymers: relaxation times and viscosities shift from scaling linearly with molecular weight for $N<N_c$ to scaling as $N^{3.4}$ for linear polymers, and even exponentially for branched polymers~\cite{watanabe1999, mcleish2002}. The tube theory, the gold standard model for understanding entangled polymer dynamics~\cite{rubinsteinPolymerPhysics2003,doi1988}, suggests that neighboring chains confine a probe chain within a tube-like region, constraining lateral motion beyond a lengthscale $a$ (the tube diameter), while allowing free diffusion along the axis of the tube (the primitive path)~\cite{doi1988}. Over recent decades, the tube theory has been enhanced to incorporate additional mechanisms such as contour length fluctuations~\cite{doi1983CLF,milnerReptationContourLengthFluctuations1998}, constraint release~\cite{rubinstein1988CR,Milner_2001} or tube dilation~\cite{marrucciRelaxationReptationTube1985, milnerParameterFreeTheoryStress1997}. This refined framework has successfully explained results from experiments on linear rheology~\cite{Likhtman2002}, dielectric relaxation~\cite{watanabeViscoelasticDielectricBehavior2004, watanabeDielectricViscoelasticRelaxation2002}, and neutron spin-echo~\cite{zamponiMolecularObservationConstraint2006}, both in equilibrium and under non-linear deformation~\cite{auhlLinearNonlinearShear2008}. Ultimately, the tube theory provides a conceptual basis for understanding how surrounding chains constrain the motion of a single polymer chain, a framework that can also be extended to active polar systems, where active forces and tube confinement yield novel dynamics.

Building on the tube model, we recently developed an analytical theory to describe the behavior of entangled polar active chains~\cite{Tejedor2019, Tejedor_2020}. For a broad range of small activities that do not disrupt chain conformations or the entanglement network, this theory predicts a linear dependence of the viscosity on molecular weight and a diffusion coefficient that is independent of the molecular weight. These predictions were confirmed through Langevin molecular dynamics simulations of active polar entangled chains diluted in a mesh of long passive entangled chains with exceedingly long relaxation times~\cite{tejedorMolecularDynamicsSimulations2023}. These simulations also naturally accounted for contour length fluctuations (CLF), \textit{i.e.} the variations in the length of a polymer chain due to thermal motion and segmental dynamics. 

\begin{figure}[hbt!]
	\centering
	\includegraphics[width=\columnwidth]{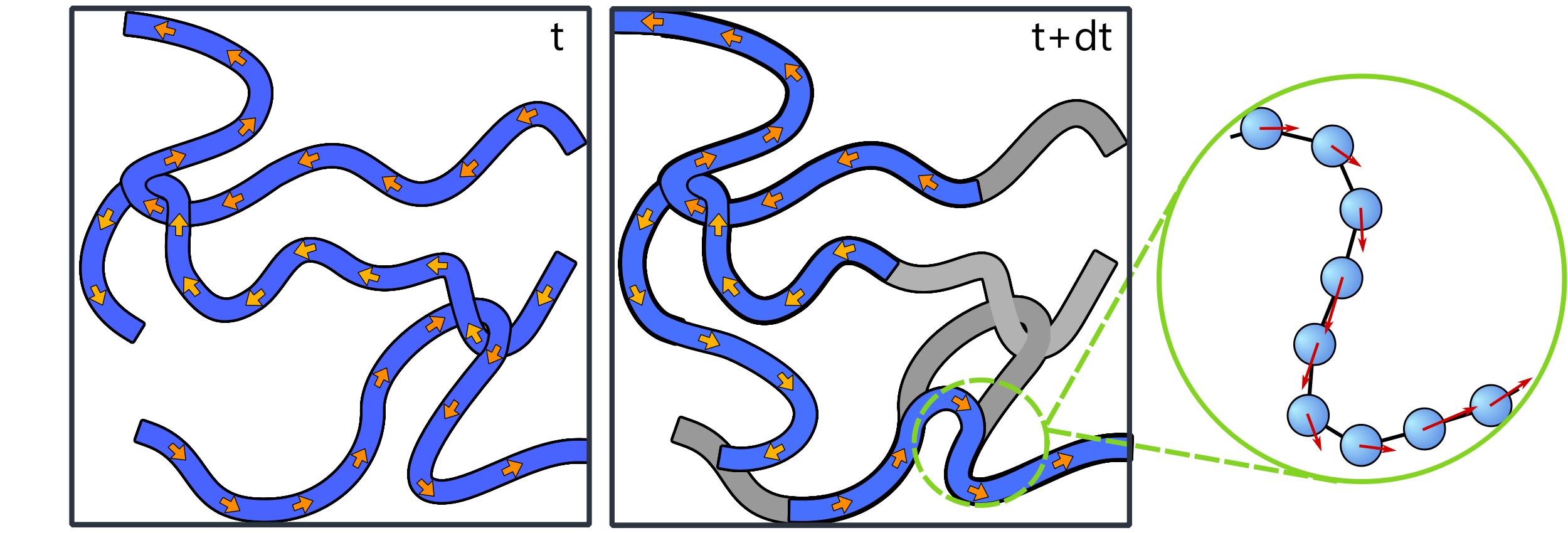}
	\caption{Schematic depiction of the polar active force applied in our coarse-grained model, and how it produces a skewed reptation motion to all chains in the system, accelerating their escape from their tubes, and affecting other multi-chain relaxation mechanisms such as constraint release (CR).}
	\label{fig: Sketch}
\end{figure}

Constraint release (CR) introduces an additional relaxation mechanism, as the confining tube around a polymer chain is not static but composed of neighboring chains that also move within their own tubes. As a result, the expected lifetime of entanglements becomes comparable to the disengagement time due to active reptation. For a given probe chain, the tube can be destroyed by reptation --- when a chain end reaches any point along the tube --- or modified by the ongoing renewal of the tubes formed by the surrounding chains with which it is entangled (see Fig.~\ref{fig: Sketch}). In a melt of polar active polymers where all chains are active, this CR effect significantly accelerates the overall system relaxation. Since CR is an intrinsically multi-body effect, multi-chain Molecular dynamics simulations provide and ideal framework for exploring its impact~\cite{Wang_2008a}.

In this work, we investigate the effect of CR on the dynamics of active polar polymers in the melt through Langevin molecular Dynamics simulations of a coarse-grained model, with polar activity uniformly applied across all polymer chains in the system. In Section~\ref{sec: methods}, we describe the coarse-grained model and the method for implementing the polar activity in our simulations. Section \ref{sec: Results} presents an analysis of the impact of polar activity on the conformation and dynamics of these polymers, with comparisons to theoretical predictions~\cite{Tejedor_2020} and results from previous studies on active diluted chains~\cite{tejedorProgressivePolymerDeformation2024} or active melts without CR~\cite{tejedorMolecularDynamicsSimulations2023}. This section concludes with a phase diagram summarizing the distinct qualitative behaviors of active polar chains as a function of P\'{e}clet and molecular weight. Finally, in Section \ref{sec:conclusion}, we highlight the key findings of this study. By extending the theoretical framework to multi-chain simulations and examining the resulting dynamics, we aim to offer new insights into active polymer systems.

\section{\label{sec: methods}Simulation method}

We carry out Langevin molecular dynamics (MD) simulations to study the dynamics and conformations of linear active entangled polymers. Polymers are modelled by the the Kremer-Grest (KG) model~\cite{Kremer_1990} ---a standard coarse-grained model for investigating the universal properties of entangled polymer melts. This model incorporates all the essential physical features, including chain connectivity, chain uncrossability (entanglements), contour-length fluctuations (breathing modes) and constraint release. The KG model describes the polymer chains as a linear sequence of purely repulsive beads connected by non-linear springs. Non-bonded interactions are modelled by the Weeks-Chandler-Andersen (WCA) potential~\cite{Weeks1971}, given by
\begin{equation}
    U_{\mathrm{WCA}} (r) = \left\{ 
    \begin{aligned}
        & 4\epsilon \left[ \left( \frac{\sigma}{r}\right)^{12} - \left(\frac{\sigma}{r}\right)^6\right]+ \epsilon, \ \ & r<r_{cut} \\
        & 0, & r \ge r_{cut} 
    \end{aligned}
    \right.
\end{equation}
where $r$ is the distance between two interacting beads, $\sigma$ the bead diameter, and $\epsilon$ the interaction strength. Neighbour beads along a polymer chain are connected by FENE springs~\cite{bird1987dynamics}, whose potential energy is given by
\begin{equation}
    U_{\mathrm{FENE}} (r) = -k \frac{R_0^2}{2} \log \left[1-\left(\frac{r}{R_0}\right)^2\right] + U_\mathrm{WCA}(r),
\end{equation}
where $k$ is the strength of the spring, and $R_0$ the maximum extensibility of the spring.

The activity is introduced as an additional force acting on each monomer, tangent to the polymer contour, as shown in Fig.~\ref{fig: Sketch}. The force on each monomer is determined by the position of its two nearest connected neighbours as:
\begin{equation}\label{eq:fa}
    \mathbf{f}^a_i = \frac{f_c}{b} \left( \mathbf{r}_{i+1}-\mathbf{r}_{i-1}\right),
\end{equation}
where $f_c$ is the magnitude of the active force and $b=0.965\sigma$ is the equilibrium value of the bond length. For the end monomers, the active force acts along the bond direction. Consequently, the active tangent force establishes a well defined direction along each polymer, allowing the ends to be identified as the \textit{head} --- the leading end --- and the \textit{tail} --- the trailing end. Discretization of the tangent force chosen in Eq.~(\ref{eq:fa}) implies that the modulus of the force acting on each monomer varies with the local chain conformation, being maximum when neighbour bonds are aligned and minimum when they adopt an anti-parallel conformation (see the right panel in Fig.~\ref{fig: Sketch}).

The evolution of the system is determined by the Langevin equations of motion for all the beads in the system, that referred to the position of particle $i$ can be expressed as
\begin{equation} \label{eq:lang}
    m\mathbf{\ddot{r}}_i = -\mathbf{\nabla}_i U(r) +\mathbf{f}_i^a -\zeta\mathbf{\dot{r}}_i + \mathbf{f}_i^r
\end{equation}
where $m$ is the mass of the bead, $U(r)= U_{\mathrm{WCA}} (r) + U_{\mathrm{FENE}} (r) $ is total potential energy, $\zeta$ is the friction coefficient, $k_B$ is Boltzmann's constant, $T$ is the temperature and $\mathbf{f}_i^r$ is a Gaussian distributed random force process~\cite{ottingerStochasticProcessesPolymeric1996} that satisfies the fluctuation-dissipation theorem~\cite{van1992stochastic}, \textit{i.e.} it has mean $\langle \mathbf{f}_i^r (t)\rangle = \mathbf{0}$ and variance $\langle f_{i\alpha}^r(t) f_{j\beta}^r(t')\rangle = 2\zeta k_B T\delta_{ij}\delta_{\alpha\beta}\delta(t-t')$, with $\alpha,\beta=x, y, z$. All simulations in this work have been run using the LAMMPS software~\cite{LAMMPS} (version 21 Jul 2020) modified to introduce the tangent active force. 
          
In the following, we use Lennard-Jones units, so that the fundamental quantities $\epsilon$, $m$, $\sigma$ and $k_B$ are all equal to 1, and the derived characteristic time $\sqrt{\sigma^2m/\epsilon}$ is taken as the unit of time. Furthermore, as in the original work of the KG model~\cite{Kremer_1990}, we use the parameters $k=30\epsilon/\sigma^2$, $R_0 = 1.5\sigma$, $b=0.97\sigma$, $\zeta=0.5$, and overall monomeric density $\rho = 0.85/\sigma^3$.
We study systems consisting of KG polymers 
with lengths $N$ varying from 50 to 800 monomers per chain. For each molecular weight $N$, the cubic simulation box side is set to ensure a minimun length of at least twice the root mean squared equilibrium end-to-end distance $\sqrt{\langle R^2_0\rangle}$, where $\langle R^2_0\rangle= C_\infty N b^2$, and the characteristic ratio $C_\infty =1.88$ as known for the Kremer-Grest model~\cite{Sliozberg_2012}. More details about the systems set up and equilibration are provided in the Supplementary Material.

To quantify the effect of activity we define the microscopic or monomeric P\'{e}clet number (Pe$_m$) defined as a the ratio of the local active and thermal forces, i.e. $\mathrm{Pe}_m= f_cb/(k_BT)$~\cite{Bianco_2018,tejedorProgressivePolymerDeformation2024,anand2018structure}. In this work, values of Pe$_m$ spanning 4 orders of magnitude are studied. Furthermore, since the motion of a polymer chain is governed by different length scales, it is useful to additionally consider a global P\'{e}clet number, defined as 
Pe$_g=N$Pe$_m$, which depends on the polymer molecular weight~\cite{tejedorProgressivePolymerDeformation2024,isele2015}.
Hydrodynamic interactions are not considered here since they are not expected to play an important role at the monomer densities studied in this work.





\section{Results and discussion}\label{sec: Results}

\subsection{Conformation and static properties}

\subsubsection{Coil average size}
\label{subsec:RG}
\begin{figure}[hbt!]
	\centering
	\includegraphics[width=\columnwidth]{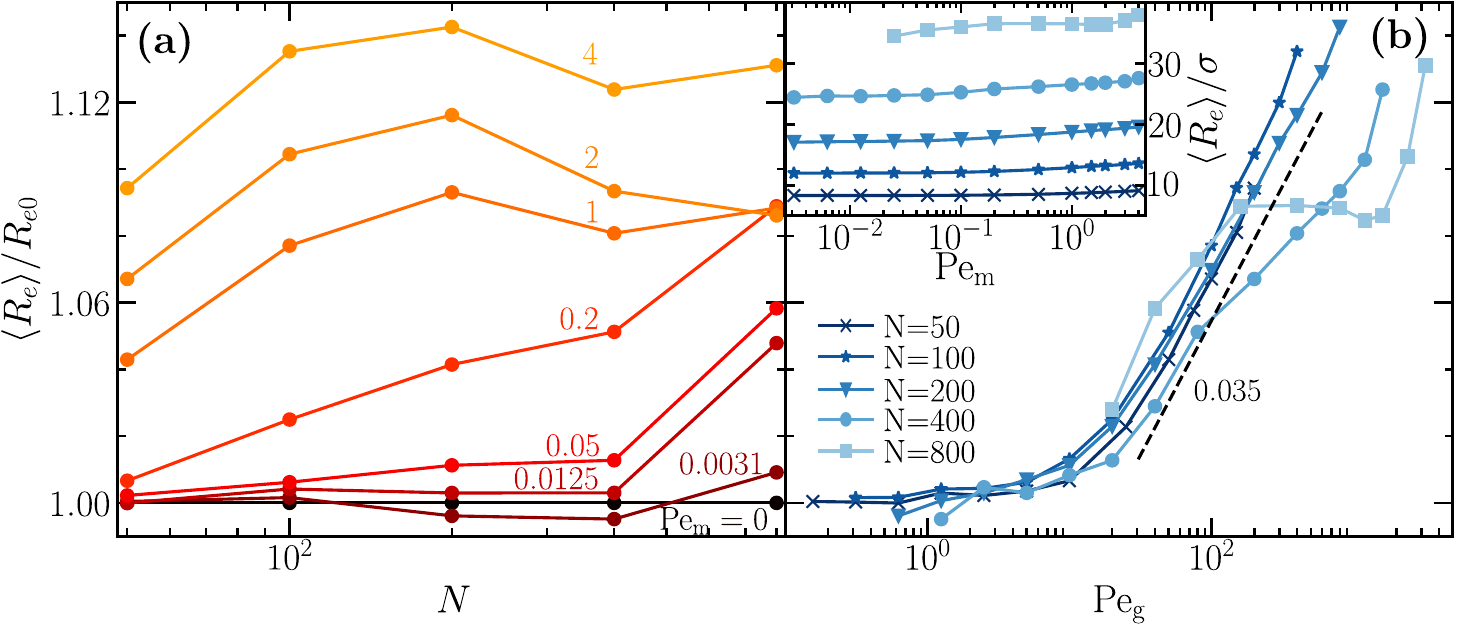}
	\caption{a)~End-to-end distance normalized with the equilibrium value, a)~as a function of the molecular weight $N$, for all studied activities; b)~as a function of the activity, for all molecular weights. 
    The dashed line is shown to highlight the logarithmic dependence of the end-to-end vector, as $\langle R_e\rangle /R_{e0} = \alpha\log(\mathrm{Pe}_g)$, with $\alpha=0.035$.
    The inset in~b) shows the absolute end-to-end distance, this is normalized with the bead diameter.}
	\label{fig: FigRe}
\end{figure}

The effect of the polar activity on the coil characteristic size is calculated by averaging the end-to-end distance of all polymers and normalizing it by the corresponding value at equilibrium $R_{e0}$ as shown in Fig.~\ref{fig: FigRe}. 
Overall, polymer conformations are only slightly stretched by activity, with a relative growth of the end-to-end distance is up to 14\% for the highest values of the activity here studied.
Qualitatively, this behavior has an opposite trend to that in the dilute regime, where both in the presence and absence of inertia, the coil size of long chains decreases by up to~30\%~\cite{tejedorProgressivePolymerDeformation2024, Bianco_2018} in a similar range of activities. 
The collapse occurring for the same chains in dilute conditions is prevented by the large confinement effects provided by the high density of monomers in the melt. Beyond $N\ge 200$,  entanglements have a significant effect and seem to slow down the growth of the coil with activity. Recently reported simulations of the same system with a higher friction coefficient report a decrease of the coil size with activity \cite{ubertiniUniversalTimeLength2024}, which indicates that inertia might affect the static properties (see discussion in Section \ref{subsec: Inertia}).


The increase of coil size as a function of activity is shown in  Fig.~\ref{fig: FigRe}(b), where the coil size shows to be almost unperturbed up to a critical value of the  P\'{e}clet number,  Pe$_g\simeq0.5$. For activities larger than this critical value, a behavior close to universal is found, similar to that observed in diluted polar active chains~\cite{tejedorProgressivePolymerDeformation2024}, including the logarithmic dependence of the coil size with the activity above the threshold Pe$_g$, as depicted by the scaling line shown in Fig.~\ref{fig: FigRe}(b). 
Since Pe$_g$ is a function of the molecular weight, this collapse implies that the critical value of the monomeric activity Pe$_m$ is smaller for longer chains. 
Long chains at high activities show a non-monotonic growth as a function of the activity, which is most likely related to the later discussed emergence of local order. Note that although measurable, the elongation is relatively small for the studied Pe$_m$ values. 



\subsubsection{Coil progressive deformation}

\begin{figure}[hbt!]
	\centering
	\includegraphics[width=\columnwidth]{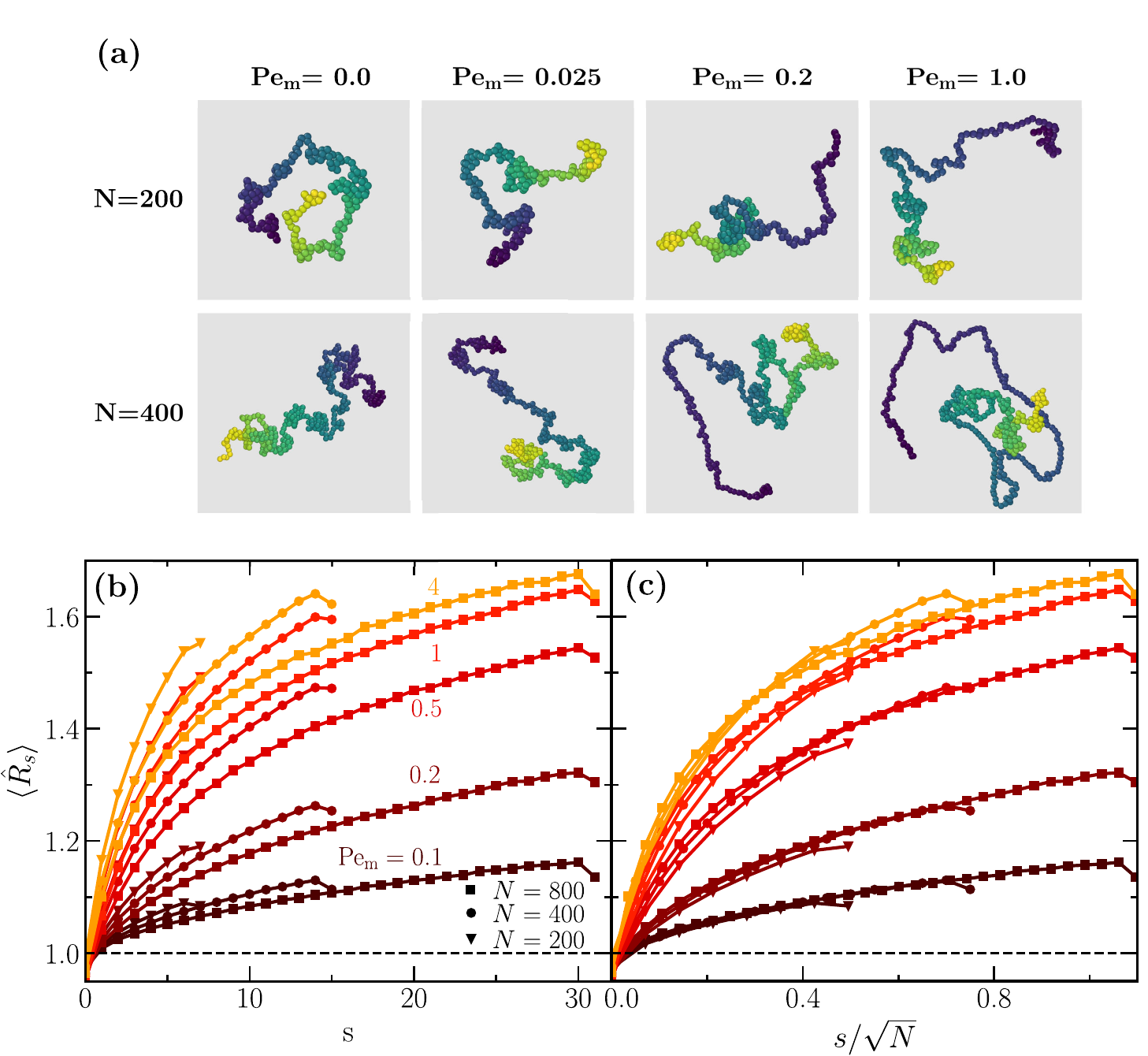}
	\caption{a)~Snapshots of  polymer conformations with $N=400$
          at various activities, with color darkness increasing
          towards the tail.
           b,c)~Segments end-to-end distance for different drifts, and
           chain lengths, b)~as a function of $s$, the distance to the
           head ($s=0$ refers to the head and $s=N/25$ to the tail);
           as a function of $s$ normalized by the square root of the
           chain molecular weight.}
	\label{fig: FigRs}
\end{figure}

While activity  only slightly stretches the overall coil size, a
visual inspection of the chain snapshots in Fig.~~\ref{fig: FigRs}(a)
reveals that the heads are significantly more compact and the tails more
elongated, similar to the dilute
case~\cite{tejedorProgressivePolymerDeformation2024}.
We quantify this effect by  $\langle \hat{R}_s\rangle$, the
mean end-to-end distance of chain segments of length $N_s=25$, which
are calculated along the chain contour and normalized by the size of strands of the same size at equilibrium. Fig.~\ref{fig: FigRs}(b) shows $\langle \hat{R}_s\rangle$ as a function of the segment position along the chain, with $s=0$
representing the head and $s=N/N_s$  the tail of the chain.
All curves show an important increase of the segments elongation towards the tail,
which already means that polar active force is breaking the self-similarity of the
polymer chains in the melt. 
Besides the very last dangling segment, the local stretching increases
with applied activity, reaching values of above 60$\%$ for the
parameters here investigated.
Increasing polymer length shows to decrease the elongation of the
segments, which is in contrast to the dilute case where the
progressive elongation depends on Pe$_m$ and is independent of $N$.
This universality was attributed to the balance between the tension transmitted and accumulated by the polar
activity along the backbone from the head to the segment located at
position $s$, and the frictional resistance of the remaining strand between $s$ and the tail.
In the case of the melt, segments at position $s$ belonging to shorter
chains elongate more than segments at the same position $s$  of longer
chains (see Fig.~\ref{fig: FigRs}(a), indicating that the tension is transmitted differently along the chain backbone when the polymer is in an active melt, due to the confinement provided by the neighboring polymers. 
This means that, given a certain segment $s$,  the tension necessary to stretch it is larger the longer the polymer, or similarly, that the stiffness of the confining network is larger the longer the building polymers are. 

Figure~\ref{fig: FigRs}(c) shows a reasonable overlap of the curves corresponding to different molecular weights onto a universal shape when the $s$ axis is rescaled by $\sqrt{N}$ .
%
To get some insight on this scaling, we recall that the end-to-end distance of a
polymer in equilibrium is proportional to $\sqrt{N}$, and this scaling is not significantly modified by the activity (see Fig.~\ref{fig: FigRe}). The fact that the same deformation is obtained for segments located at position $s/\sqrt{N}$ in chains of different lengths suggests that the same average force-drag balance is obtained when the relative location of the projected position of segment $s$ onto the end-to-end vector is also the same (see a sketch in Fig.~S1(a) in the Supplementary Material). 

\subsubsection{Mesh architecture}\label{subsec: PPA}

In the context of reptation, it is crucial to test whether activity modifies the underlying entanglement network. For that purpose, we performed a primitive path analysis (PPA) \cite{sukumaran2005, everaers2004} on steady-state snapshots of all simulated systems, using a modified version of LAMMPS~\cite{hagitaEffectChainpenetrationRing2021}. 
From this analysis, we can extract the average length of the primitive path $L_\mathrm{PP}$ and, assuming random walk statistics, relate it to the diameter of the tube as, $a=\langle R^2 \rangle /L_\mathrm{PP}$, where $\langle R^2 \rangle$ is the mean-square end-to-end distance of the chains. 
\begin{figure}[hbt!]
	\centering
	\includegraphics[width=\columnwidth]{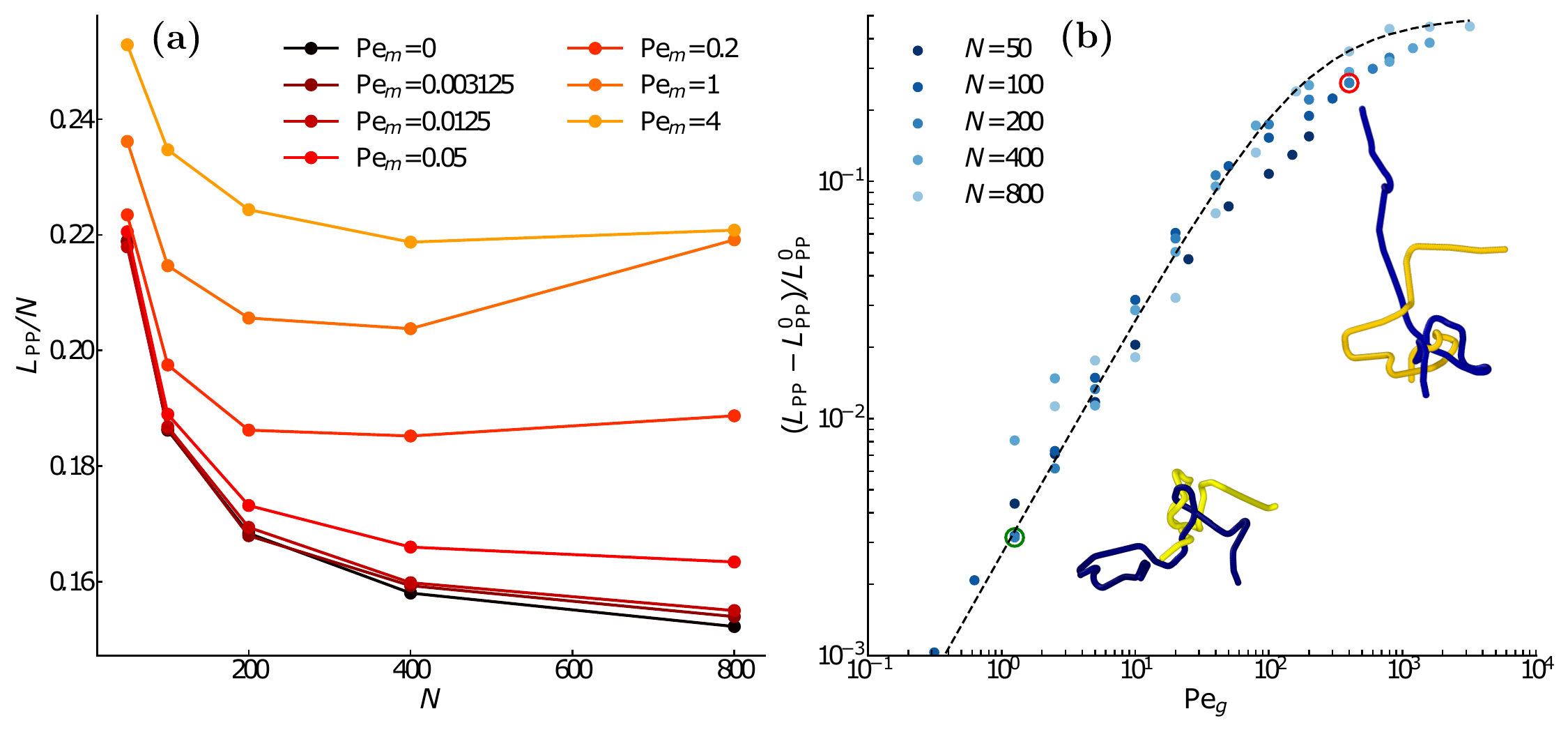}
	\caption{ 
 a)~Normalized average tube length $L_\mathrm{PP}$, calculated from Primitive Path Analysis (PPA), as a function of the molecular weight $N$, for different values of the activity, showing the stretching of the tubes. 
 b)~Relative growth of the primitive path length as a function of the
 global P\'eclet number, Pe$_g$ = Pe$_mN$. Dashed line corresponds to
 the universal behavior of Eq.~(\ref{eq: fit}). Insets: snapshots of two representative pairs of entangled primitive paths from chains of $N=200$. Bottom snapshot corresponds to the green highlighted point with a Pe$_m$=0.00625, and top snapshot to the red highlighted point with Pe$_m$=2 , showing signs of tube elongation due to activity.}
	\label{fig: FigPPA}
\end{figure}

In Fig.~\ref{fig: FigPPA}(a), the average length of the primitive path $L_\mathrm{PP}$, normalized by the molecular weight $N$, is plotted versus the molecular weight for different values of activity.
For reference, the value of the polymer melt in equilibrium is also displayed, showing the usual overestimation of the tube length by PPA when the polymer chains are short. 
For small values of the activity, Pe$_m\leq 0.0125$, the tube length is almost indistinguishable with respect to the equilibrium primitive path.
With increasing activities, the primitive path gets progressively more elongated. In Fig.~S2 of the Supplementary Material, example snapshots of primitive path networks after PPA are represented, showing clear signs of tube elongation at high activities.

To better understand the increase in tube length with activity, Fig.~\ref{fig: FigPPA}(b) shows the relative increase in the length of the primitive path with respect to the obtained equilibrium length $L_\mathrm{PP}^0$, plotted as a function of the global P\'{e}clet Pe$_g=$Pe$_mN$. The data reveals a seemingly universal linear increase at low to moderate activities, followed by saturation at very high values of Pe$_g$.
In the inset of Fig.~\ref{fig: FigPPA}(b), two snapshots of a representative primitive path configurations  are shown at low and high activities, revealing the tube elongation due to the activity.
This data also shows that the primitive paths extend up to 50\% of their size in equilibrium. This is much more that the  average elongation of the end-to-end distance of the individual polymers which increase a maximum of 10\% for the ranges here investigated (see Fig.~\ref{fig: FigRe}). 
This difference suggests that the chains are forming an inward-folded structure, as evidenced by the snapshots in Fig.~\ref{fig: FigRs}.

In order to provide an estimation of the length of the tube we assume that it results from balancing the elongation force of the active term, $f_{a}$, with the elastic retraction of the polymer confined within the tube,  $f_{e}$.
The total active tension accumulated in the chain is 
\begin{equation}
    f_\mathrm{a}\propto N f_c, \label{eq:fact} 
\end{equation} and tends to stretch the chain inside the tube.
The resistance to such stretching comes from the entropic elasticity of the polymer chain within the tube, which can be assumed to have a similar nature to the FENE force that governs the bonded forces, that is, 
\begin{equation}
    f_{e}=k_e\frac{L_\mathrm{PP}-L_\mathrm{PP}^0}{1-\left(\frac{L_\mathrm{PP}}{L_\mathrm{PP}^\mathrm{max}}\right)^2}. \label{eq:felas}
\end{equation}
Here $k_e=3k_BT/Nb^2$ is the elastic constant, and $L_\mathrm{PP}^0$ the equilibrium contour length of the primitive path which, according to the tube theory, is $L_\mathrm{PP}^0=Nb^2/a$, being $a$  the tube diameter, and $L_\mathrm{PP}^\mathrm{max}$ is the maximum length of the tube, which we take here as $L_\mathrm{PP}^\mathrm{max}\simeq 1.5 L_\mathrm{PP}^0$.
Considering now that  $f_{a}=f_{e}$ we have that
\begin{equation}
    N f_c = \frac{3k_BT}{Nb^2} \frac{L_\mathrm{PP}-L_\mathrm{PP}^0}{1-\left(\frac{L_\mathrm{PP}}{L_\mathrm{PP}^\mathrm{max}}\right)^2}
\end{equation}
Dividing on both sides by $L_\mathrm{PP}^0$, replacing $f_c$ with the expression for P\'{e}clet number and simplifying we get:
\begin{equation}
    \frac{L_\mathrm{PP}-L_\mathrm{PP}^0}{L_\mathrm{PP}^0} = \frac{N \mathrm{Pe}_m a}{3b}\left[1-\left(\frac{L_\mathrm{PP}}{L_\mathrm{PP}^\mathrm{max}}\right)^2\right]
\end{equation}
which shows that the relative growth of the tube length $L_\mathrm{PP}$ depends only on the global P\'{e}clet number $\mathrm{Pe}_g = N \mathrm{Pe}_m$.
Active tension acts along the contour of the chain and not on the primitive path of the tube.
The projection of the active tension on the tube axis is expected to be much smaller than the one provided along the polymer contour by $f_a$, such that we can consider a fit parameter $\alpha$ resulting into 
\begin{equation}
    \frac{L_\mathrm{PP}-L_\mathrm{PP}^0}{L_\mathrm{PP}^0} = \alpha \mathrm{Pe}_g\left[1-\left(\frac{L_\mathrm{PP}}{1.5 L_\mathrm{PP}^0}\right)^2\right]
    \label{eq: fit}
\end{equation}
The black dashed line shown on Fig.~\ref{fig: FigPPA}(b) has been obtained with $\alpha=0.00485$. More details about the derivation of Eq. \eqref{eq: fit} 
can be found in Section SIII of the Supplementary Material.


\subsubsection{Limits of validity of the active reptation theory}

A theory to predict the influence of polar active in the reptation properties of a polymer melt has been recently developed in the limit of small activities~\cite{Tejedor2019, Tejedor_2020}. This theory builds upon a basic tube theory assumption, that the chain ends can explore all possible orientations before creating new tube segments, so that the tube maintains the fractal structure of a random walk. This assumption implies that the maximum drift velocity of the primitive path along the tube $c_{max}$ needs to smaller than the ratio of the tube diameter $a$ and the entanglement time $\tau_e$, as  $a/\tau_e$. This means that the chain end can relax its orientation (in a time of the order of $\tau_e$) before the primitive path moves by the activity-induced drift a distance of the order of the tube diameter $a$. 
An estimation of $c_{\max}$ considers the values obtained for the standard KG model, $\tau_e \approx 5800$ and $a\approx 9.1$~\cite{likhtman_2007}, resulting into $c_{\max}\approx 1.57\cdot 10^{-3}$, which is very small.  
In addition, constraint release is not considered by the theory, so tubes are described as static objects, an assumption which is expected to breakdown when all chains are active, as explained already in Fig.~\ref{fig: Sketch}. If the tube is static, the correlation of the tube segment orientation is expected to simply decay exponentially with distance.

\begin{figure}[hbt!]
	\centering
	\includegraphics[width=0.7\columnwidth]{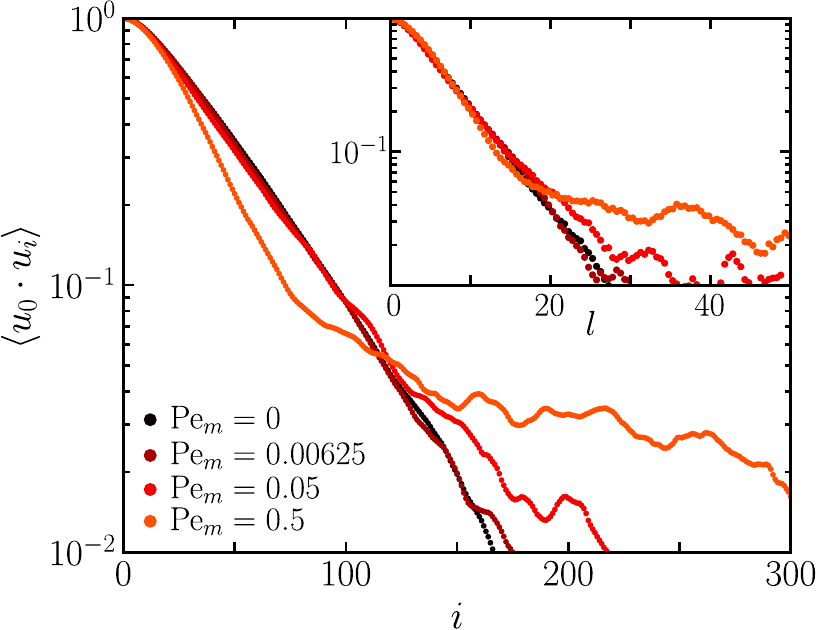}
	\caption{Orientation correlations of bonds along the primitive path from head to tail, $\langle {\bf u}_0 \cdot {\bf u}_i \rangle$, for chains with molecular weight $N=400$ and different activities, as a function of monomeric distance to the head along the primitive path, $i$. 
    Inset:  $\langle {\bf u}_0 \cdot {\bf u}_i \rangle$  as a function of the contour length distance along the primitive path, $l$.}
	\label{fig: Figbondcorr2}
\end{figure}
To precisely quantify this effect, we here consider the unit vector of the bonds along the primitive path, $u_i$, and calculate their mean correlation. In Fig.~\ref{fig: Figbondcorr2}, $\langle {\bf u}_0 \cdot {\bf u}_i \rangle$ is shown as a function of $i$, the monomeric distance (as obtained from PPA) from the head along the primitive path, for different activities. The passive case shows an exponential decay with $i\approx 40$ monomers. This number can be compared with the typical number of monomers in between entanglements, $N_e$, which for the Kremer-Grest model is $N_e=52$, as determined by Likhtman from the plateau modulus~\cite{likhtman_2007}. 
At very small activities, Pe$_m\le 0.01$, the correlation is almost identical to the equilibrium case, indicating that the orientational properties of the entanglement network are not being perturbed by the activity. 
However, at slightly larger activities Pe$_m\ge 0.05$, the correlation deviates from the equilibrium decay at monomeric distances above $i>100$, which is a sign that orientational correlation between tube segments is appearing beyond the entanglement molecular weight. This already indicates that the main assumption in the active reptation theory at these still relatively small activities already fails.
For a bit larger activities (Pe$_m=0.5$), a stronger decay appears at very small monomeric distances $i<N_e$, which does not occur when plotting the orientation correlation as a function of the contour length distance along the primitive path (see inset in Fig.~\ref{fig: Figbondcorr2}). This suggests that the tube remains unperturbed next to the head, while the chain inside is getting stretched (see Fig. S1(b) in the Supplementary Material), and that the correlation very clearly deviates from the exponentially one at distances $l$ beyond the equilibrium tube diameter $a$. 

Although CR effects were not included in the active reptation theory~\cite{Tejedor2019,Tejedor_2020}, some predictions remain valid for low activity levels. However, for activities above Pe$_m\ge 0.05$, we expect deviations, given the elongation and orientational correlation observed in tube segments. In molecular dynamics simulations of active chains moving within a network of long passive polymers~\cite{tejedorMolecularDynamicsSimulations2023}, where CR was intentionally disabled, no tube deformation was observed up to Pe$_m=0.05$. This suggests that CR, by softening the tubes, amplifies the impact of activity on both the chain conformations and the overall entanglement network.

\subsubsection{Activity-induced bond alignment}\label{subsec: OrderParameter}

When  polymers are subjected an external flow, 
chain segments typically orient in the flow direction. In melts, this orientation occurs when the flow rate exceeds the inverse of the disengagement time $\tau_d$, and before any signs of elongation appear, which arises only when the flow rate exceeds the inverse of the Rouse time $\tau_R$~\cite{grahamMicroscopicTheoryLinear2003}. 
However, in polar active polymer melts, no flow field is present, so there is no preferred orientation, and the overall orientation distribution of remains isotropic. 
Nonetheless, it is not immediately clear whether polar activity might still lead to some degree of local alignment. 


We calculate then the order parameter $p_2$ to determine whether alignment precedes, or drives chain stretching.
To quantify the degree of alignment and orientation of the polymer segments, we first compute, for each monomer $i$, the unit vector pointing from the previous monomer to the next along the chain contour. Then, we calculate the angle $\theta_{ij}$ between this vector and the unit vectors associated with all monomers $j$ surrounding monomer $i$. Finally, we determine the second-order Legendre polynomial parameter $p_2^i$ for each monomer (except for the ends), defined as:
\begin{equation}\label{eq:p2Legendre}
   p_2^i = \left\langle \frac{3 \cos^2 \theta_{ij} - 1}{2} \right\rangle_j,  
\end{equation}
where the average runs over all monomers $j$ within a cut-off distance $r_c$ of monomer $i$ (in this case, $r_c = 3.0 \sigma$). The order parameter $p_2$ is commonly used to detect nematic alignment with $p_2 = 1 $ indicating perfect alignment and  $p_2 = 0$ a random distribution of orientations. In our simulations, $p_2^i$ is computed for each monomer and  $\langle p_2\rangle$ is averaged over monomers and steady state snapshots.



\begin{figure}[hbt!]
	\centering
	\includegraphics[width=\columnwidth]{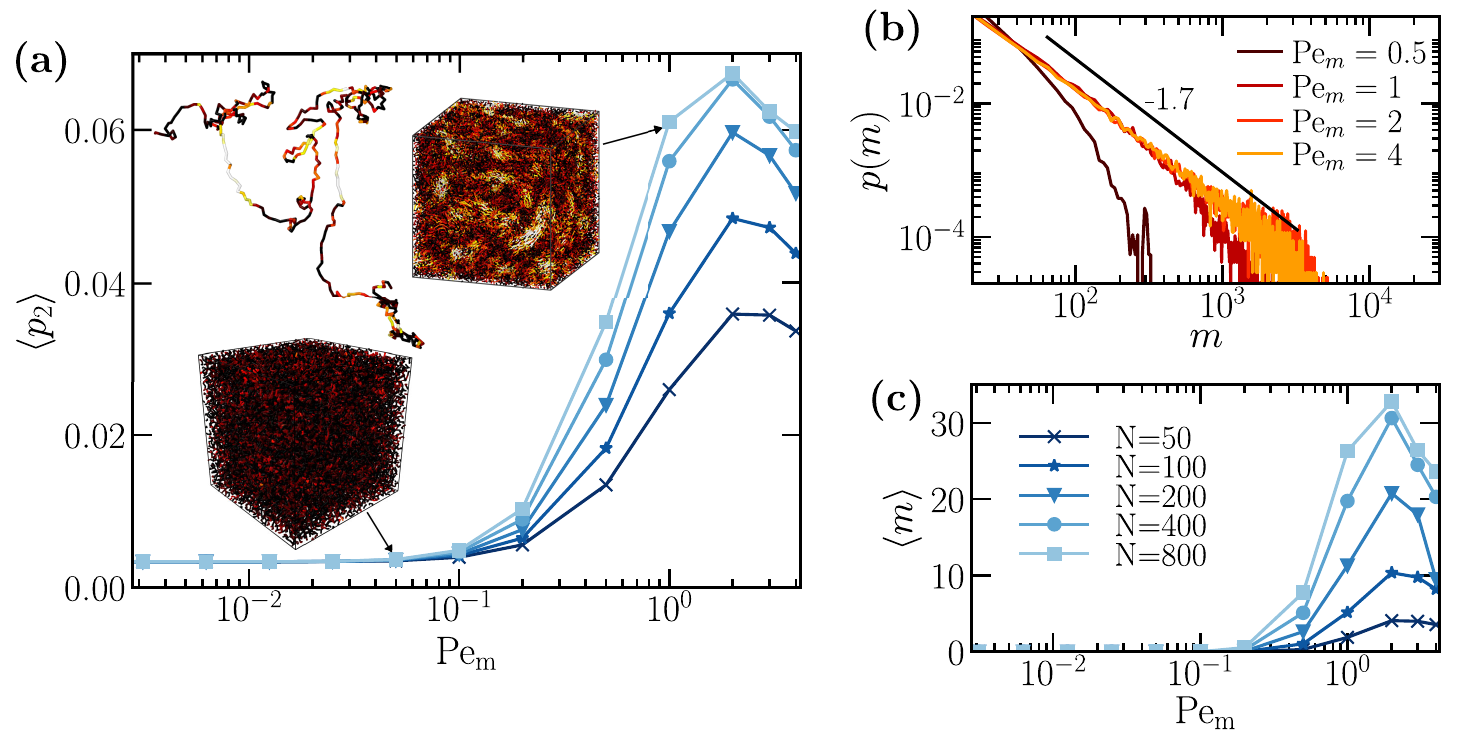}
	\caption{
    a)~Average order parameter, $\langle p_2 \rangle$ as a function of $\text{Pe}_m$ for the molecular weights studied. Snapshots for $N = 800$ are included as insets, with atoms colored from black (low $p_2$) to red, white and yellow (higher $p_2$). Two snapshots refer to the bond distribution of the complete simulated melt for low and high indicated activity. For the high activity case the third snapshot illustrates the different degrees of orientation along a single chain contour.  
    b)~Probability distribution of nematically ordered monomer clusters for different activities in systems of $N=200$. 
    c)~Mean cluster size as a function of Pe$_m$ for all molecular weights.}
	\label{fig: Figp2}
\end{figure}
In Fig.~\ref{fig: Figp2}(a), the mean value of the order parameter $\langle p_2\rangle$ is shown as a function of Pe$_m$, for different molecular weights. 
In equilibrium, $\langle p_2\rangle$ has a non-vanishing value related to some local alignment due to excluded volume effects. Increasing the activity level to Pe$_m=0.1$ does not result in any noticeable increase in $\langle p_2\rangle$. However, beyond this threshold, activity clearly makes local bond alignment to increase, effect which is stronger for longer polymers. All curves exhibit a peak alignment at Pe$_m\approx 2$, with the maximum value increasing with molecular weight, followed by decrease at higher Pe$_m$ values. 
The mean values of $\langle p_2\rangle$ remain relatively low due to the inhomogeneous bond alignment within the simulation box (see snapshots in Fig.~\ref{fig: Figp2}(a). At low $\text{Pe}_m$, the simulation box appears mostly dark, indicating a lack of nematic order. At higher activities, bright regions emerge, indicating local clusters of aligned segments. 
The snapshot of a single molecule nicely illustrates how $p^i_2$ varies along the chain contour, and that straight segments more frequently correspond to higher $p_2$
As with segmental stretching (see Fig.~\ref{fig: FigRs}), nematic order is higher in segments closer to the tail (see Fig.~S4 in the Supplementary Material).

The clusters of aligned bonds are reminiscent of nematic tactoids~\cite{lettinga2021tactoids,schoot2022tactoids}, and in order to characterize their average size $m$ we first define a threshold value $p_2^{th} = 0.25$, above which a monomer is considered to be ordered. We then identify clusters of ordered monomers based on a distance criterion: ordered monomers within a maximum distance of $d=2.0\sigma$ are considered part of the same cluster. 
Figs.~\ref{fig: Figp2}(b) and~\ref{fig: Figp2}(c) show the cluster size distribution $p(m)$ and mean value of the size $\langle m \rangle$ of the clusters of aligned bonds, respectively. 
Below a threshold of Pe$_m<1$,  $p(m)$ decays exponentially, broadening as activity increases. At higher activities, the distribution adopts a power-law shape  $p(m) \sim m^\nu$ with  $\nu\simeq -1.7$, resembling scale-free networks. Similar cluster size distributions have been observed in other active matter models, such as the Vicsek model~\cite{huepeIntermittencyClusteringSystem2004}, self-propelled hard disks~\cite{levisClusteringHeterogeneousDynamics2014}, self-propelled rods~\cite{peruaniNonequilibriumClusteringSelfpropelled2006}, and active Brownian particles with polar alignment~\cite{martin-gomezCollectiveMotionActive2018}. These distributions typically precede percolation and eventually also the emergence of a giant cluster. 

By closely inspecting the curves in Fig.~\ref{fig: Figp2}(c), a maximum cluster size is observed at Pe$_m=2$, followed by a slight decrease at Pe$_m=4$. This peak in cluster size aligns with the maximum values of $\langle p_2\rangle$ (Fig.~\ref{fig: Figp2}(a)), suggesting that activity promotes bond alignment and cluster growth up to a threshold Pe$_m=2$. Beyond this point, higher activity disrupts further alignment and the formation of larger clusters. Interestingly, polar activity still generates clusters of several thousand aligned bonds (Fig.~\ref{fig: Figp2}(b), with longer chains forming larger clusters (Fig.~\ref{fig: Figp2}(c)). However, these clusters exhibit highly dynamic behavior, continuously dissolving and reforming at distant, seemingly uncorrelated regions of the simulation box. This rapid fluctuation prevents the formation of distinct ordered and disordered regions, and no phase separation is observed in our simulations.

\subsection{Dynamical properties}

As in the case of dilute polymers with polar activity, the total active force acting on each polymer center of mass is proportional to its end-to-end vector. However, in a melt, chains are constrained and must slither along their primitive paths rather than move freely in the direction of the force, as they do in dilute conditions. Despite this constraint, activity is expected to significantly impact the dynamical properties of the melt. 
To investigate this, we examine the center-of-mass and segmental mean-square displacement, end-to-end vector relaxation, and tangent-tangent correlation function. All these dynamical observables have been calculated using an efficient multiple tau correlator technique~\cite{Ramirez_2010}. 

\subsubsection{Center of mass MSD and diffusion coefficient}\label{subsec: MSD of COM}

\begin{figure}[hbt!]
	\centering
	\includegraphics[width=\columnwidth]{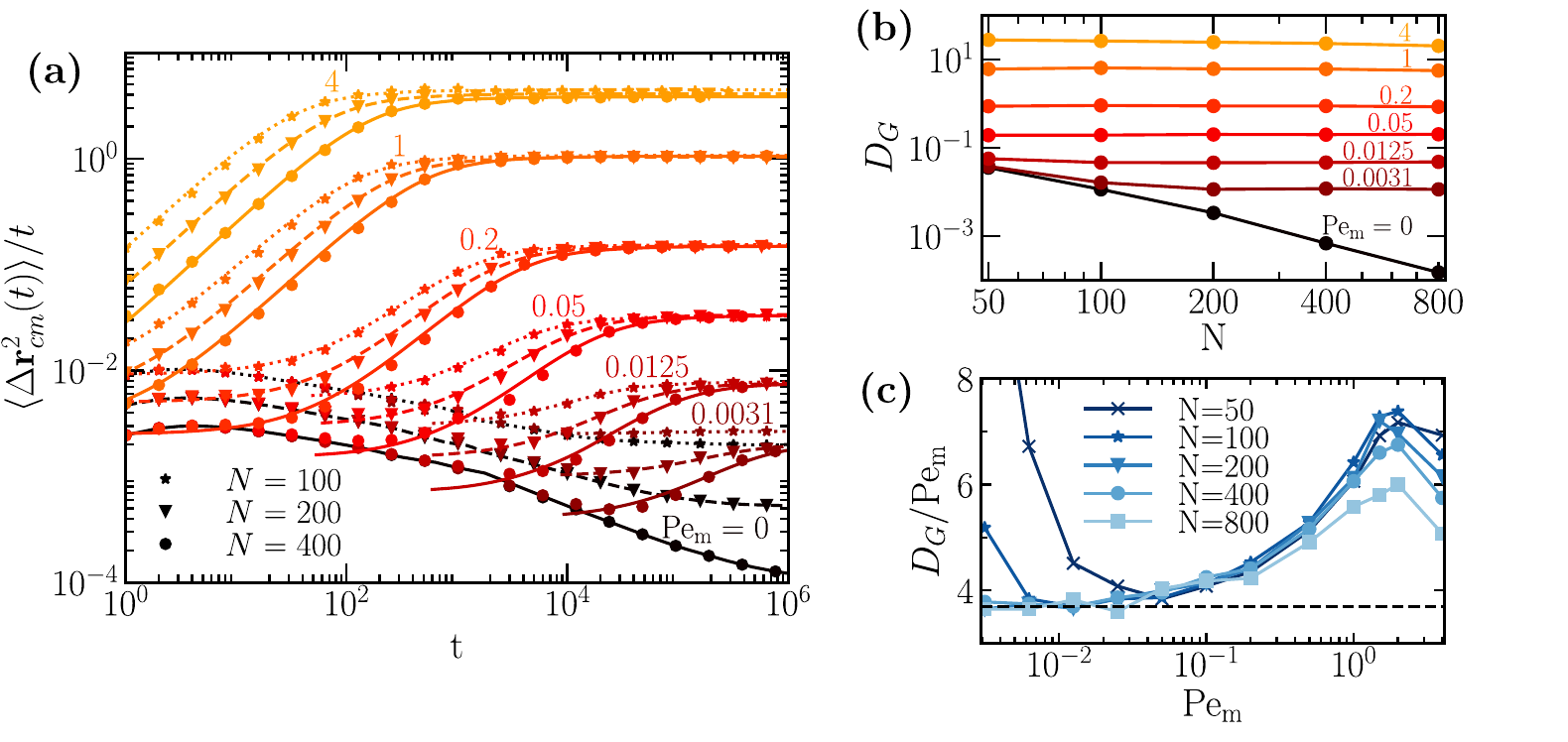}
	\caption{a)~Molecular center of mass mean square displacement normalized by time for chains of various lengths and applied polar activities, showing superdiffusive motion at short times and diffusive at long times. Lines correspond to the fit to the ABP behavior (see Supplementary Material).
    b)~Diffusion coefficient vs molecular weight for each value of the active force. 
    c)~Diffusion coefficient, normalized by Pe$_m$ vs. Pe$_m$ for all molecular weights. 
    Dashed line is a guide to eye showing the expected scaling from the active reptation theory~\cite{Tejedor2019}.}
	\label{fig: FigDiffAndg3}
\end{figure}

The mean square displacements (MSD) of the centers of mass of chains with different molecular weights at various Pe$_m$ are presented in Fig.~\ref{fig: FigDiffAndg3}(a). 
In the passive case, the expected long subdiffusive behavior is observed, which at longer times  transitions into a molecular weight-dependent Fickian regime at the disengagement time $\tau_d$~\cite{doi1988}, appearing as a plateau with the chosen normalization. In the case of activity governing the dynamics over reptation, two key phenomena emerge: (i) the terminal Fickian regime, observed at times $t>\tau_c$, where $\tau_c$ is the disengagement time due to activity, is preceded by a superdiffusive regime, which can persist for over a decade in logarithmic time at high activities, and (ii) the diffusivity becomes independent of molecular weight, depending solely on Pe$_m$. Both observations align with the active reptation theory predictions~\cite{Tejedor2019}, suggesting that CR does not significantly influence the qualitative behavior of the MSD of the center of mass.

Interestingly, when the activity dominates the dynamics, the diffusion of the center of mass can be accurately described by the MSD of an active Brownian particle (ABP)~\cite{Howse_2007}. 
The fits to the simulation results are shown as lines in Fig.~\ref{fig: FigDiffAndg3}(a) for Pe$_m>0$ (details provided in the Supplementary Material). 
The fits do not reproduce the behavior at very short times, since the equation for an ABP does not consider any of the internal degrees of freedom that are characteristic of the motion of polymer chains, but work very well from times $t$ almost two decades smaller than  $\tau_c$. 
A similar near-quantitative agreement was found in the MSD of dilute active polar polymers~\cite{tejedorProgressivePolymerDeformation2024}, which is attributed to the total active force on the center of mass being proportional to the end-to-end vector, whose orientation relaxes almost as a single exponential. 


From the Fickian regime in Fig.~\ref{fig: FigDiffAndg3}(a), the self-diffusion constant can be extracted as $D_G=\lim_{t\to\infty} \Delta \mathbf{r}_{cm}^2(t)/6t$, as shown in Fig.~\ref{fig: FigDiffAndg3}(b), where it can again be confirmed that $D_G$ becomes independent of molecular weight when the activity dominates the dynamics and  depends only on Pe$_m$.
In the passive case, $D_G$ follows a power-law dependence on molecular weight, scaling roughly as $D_G\propto N^{-1.8}$ for $N<200$ (with deviations from the expected Rouse scaling of $D_G\propto N^{-1}$ already reported in unentangled melts~\cite{liDynamicsLongEntangled2021}), and transitions to $D_G\propto N^{-2.2}$ for $N>200$, which is characteristic of reptation with contour-length fluctuations~\cite{gellViscoelasticitySelfdiffusionMelts1997a}. Since $N_e\approx 52$ in the KG model~\cite{likhtman_2007}, the transition from Rouse to reptation regimes occurs around $N\approx 3$ to $4 N_e$. 

To gain a deeper understanding of the diffusion behavior, Fig.~\ref{fig: FigDiffAndg3}(c) depicts $D_G/\mathrm{Pe}_m$, revealing three distinct regimes. 
At very low activities, diffusive reptation dominates with a constant diffusion, which makes  $D_G/\mathrm{Pe}_m$ diverge. This can be seen in  Fig.~\ref{fig: FigDiffAndg3}(c)
for shorter chains ($N=50, 100$). 
The threshold Pe$_m^t$, above which active forces overtake reptation, is expected to scale as Pe$_m^t\propto N^{-2}$~\cite{Tejedor2019}, implying that longer polymers will also show such divergence at even smaller values of the activity. 
For intermediate values of the activity, below a threshold Pe$_m\approx 0.05$, polar activity dominates over the diffusive reptation motion, and the diffusion coefficient shows to be proportional to the P\'{e}clet number, particularly for entangled polymers ($N\gtrsim 3N_e$). 
In this regime, the total active force applied along the chain contour in Eq.~(\ref{eq:fact}) increases with molecular weight as $f_a \propto N\mathrm{Pe}_m$. 
Simultaneously, the friction coefficient for the chain slithering motion along the primitive path increases as $\zeta_N=N\zeta_0$, where $\zeta_0$ is the monomeric friction coefficient, not necessarily equal to the friction coefficient in the Langevin Eq.\eqref{eq:lang} (see Section \ref{subsec: Inertia}). 
The drift velocity of the chain in its slithering motion is therefore independent of $N$, since $c\approx f_a/\zeta_N \approx \mathrm{Pe}_m/\zeta_0$. Since the length the chain must travel to escape the tube grows linearly with $N$ (i.e. $L_\mathrm{PP} \propto N$), the time required for the chain to escape the tube due to active slithering motion scales as $\tau_c = L_\mathrm{PP}/c \propto N/c$. Assuming that active chains approximately maintain Gaussian statistics, each time a chain escapes its tube, it must have traveled a distance on the order of the mean squared end-to-end distance, $\langle R^2\rangle = Nb^2$. Therefore, the diffusion coefficient can be estimated as $D_G\approx \langle R^2\rangle/\tau_c \approx bc\propto \mathrm{Pe}_m$, which agrees with simulation results shown in Fig. \ref{fig: FigDiffAndg3}(c). 

Finally, at larger values of the activity,  $D_G$ grows faster than linearly with respect to Pe$_m$, reaching a peak at Pe$_m=2$, before decreasing again. This enhanced diffusion correlates with the emergence of nematic bond alignment, which also peaks at Pe$_m=2$, as shown in Fig.~\ref{fig: Figp2}. Alignment is likely to reduce the local friction coefficient for monomer sliding, thereby enhancing the overall chain motion along the tube. Interestingly, though alignment is more pronounced in longer chains, the diffusivity increase  is less pronounced. 
This may result from the combination of reduced friction and increased tube stretching, which is more significant for longer chains, as seen in Fig.~\ref{fig: FigPPA}(b). Consequently, while friction decreases due to alignment, longer chains must still reptate over greater distances under strong activity. As Pe$_m$ increases further, alignment decreases, but tube stretching persists, leading to a decline in the diffusion coefficient $D_G$. 

\subsubsection{Monomeric MSD}\label{subsec: MonomericMSD}

\begin{figure}[hbt!]
	\centering
	\includegraphics[width=\columnwidth]{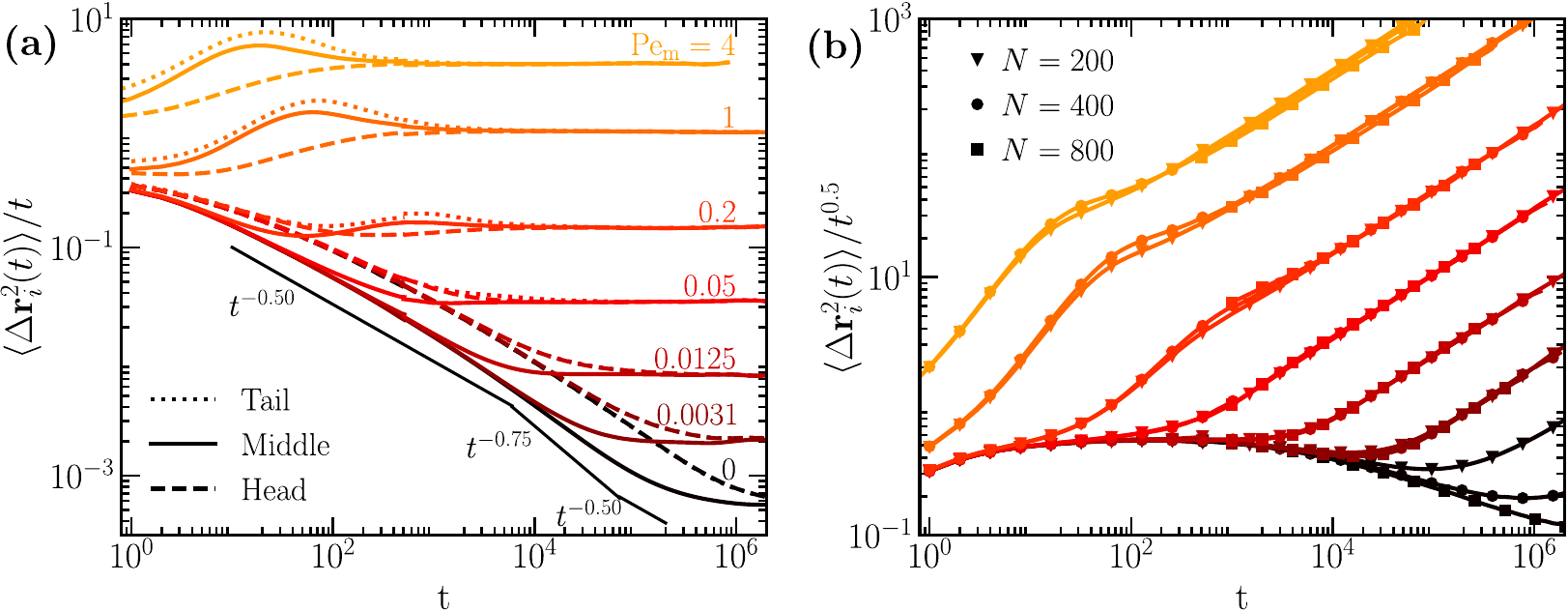}
	\caption{(a) Mean-square displacement divided by time (to highlight the terminal Fickian regime) of selected monomers --- middle bead (solid lines), head (dashed lines), and tail (dotted lines) --- for chains of molecular weight $N=200$ and  different values of Pe$_m$. Thin, black solid lines highlight the characteristic power laws predicted by the tube theory~\cite{doi1988}. (b) MSD of the central monomer normalized by $t^{1/2}$ (to highlight the Rouse scaling) for the same activities as in panel (a) and molecular weights $N=200$ (triangles) and $N=400$ (circles). }
	\label{fig: Figg1}
\end{figure}

In Fig.~\ref{fig: Figg1}(a), the time-normalized MSD for selected monomers of chains with $N=200$ at various Pe$_m$ values are shown. 
In the passive case, the MSD follows the classical power laws predicted by Doi and Edwards for a Rouse chain reptating inside a random-walk-like tube~\cite{doi1988}, with the middle monomer moving the slowest, while the head and tail, having identical MSD, move the fastest. Given that the activity disrupts the conformational symmetry between the head and tail of the chain, it is interesting to explore whether this symmetry breaking also affects their dynamics. It is worth noting that, according to the analytical theory of active reptation~\cite{Tejedor_2020} and previous simulations that include consider CLF~\cite{Tejedor_2020,tejedorMolecularDynamicsSimulations2023}, even though the activity induces a clear polarity along the chain, the dynamical symmetry between head and tail in terms of monomeric MSD is preserved. The same symmetry holds for the melts of polar active chains studied here, up to a moderate P\'{e}clet number, Pe$_m=0.0125$ (see Fig.~\ref{fig: Figg1}(a). 

However, for higher activities, a clear dynamical asymmetry emerges, with the head monomer moving more slowly than the tail. This asymmetry becomes more pronounced with increasing Pe$_m$, and for Pe$_m\geq 1$, the head monomer becomes the slowest along the entire chain, even slower than the middle monomer. These dynamics are reminiscent of those observed in active polar diluted chains~\cite{tejedorProgressivePolymerDeformation2024}, where the asymmetry was attributed to the tail loosely following the head, diffusing over a longer path, and exhibiting lateral fluctuations that increase with activity.

In the current study, however, since the chains are in a dense, entangled melt, they are constrained to move along a tube. To accommodate the greater lateral fluctuations of the tail, properties of the tube segments. such as  step length and radius, must change between the time the head creates the tube segment and the moment the tail passes through and destroys it. This effect may be driven by CR, potentially inducing an activity-enhanced dynamic tube dilation effect~\cite{vanruymbekeEffectiveValueDynamic2012} during the lifetime of each tube segment. 

Interestingly, at very high activities (Pe$_m>0.5$), all monomers exhibit a superdiffusive regime in their MSD. This phenomenon is not predicted by the analytical theory of active reptation, as the Pe$_m$ values required to observe superdiffusion fall outside the validity range of the theory \cite{Tejedor_2020}.

In Fig.~\ref{fig: Figg1}(b), the MSDs of the middle monomer for chains of different molecular weights are compared. The MSD is divided by $t^{0.5}$ to highlight two of the classical power-law regimes predicted by Doi and Edwards~\cite{doi1988}. In the passive case, the motion of the middle monomer of chains of $N=200, 400$ and 800 is nearly indistinguishable, up until  disengagement time, which depends on the molecular weight $N$. Interestingly, when the activity dominates the dynamics, the MSDs of the middle monomer for all molecular weights nearly overlap across the entire time range. As Pe$_m$ increases, the chain can escape the tube at a time $\tau_c$, which may be shorter than the disengagement time $\tau_d$, the Rouse time of the chain $\tau_R$ (both of which depend on molecular weight), or the Rouse time of an  entanglement $\tau_e$ (which is independent of molecular weight). Depending on the the strength of the activity, this may causes the loss of the corresponding power-law regimes predicted by the tube theory.  
As long as the $t^{0.25}$ power-law regime is observed in the monomeric MSD, the tube is successfully imposing a lateral constraint on the free three-dimensional motion of the chain. However, at high activity levels (Pe$_m\ge 0.05$) this characteristic power-law disappears, as the chain escapes the tube before $\tau_e$. In this case, the tube no longer fully restricts the motion of the chain, and key assumptions of the tube theory, such as the isotropic orientation of new tube segments at the ends, no longer hold~\cite{Tejedor2019}.

\subsubsection{End-to-end relaxation}
\begin{figure}[hbt!]
	\centering
	\includegraphics[width=\columnwidth]{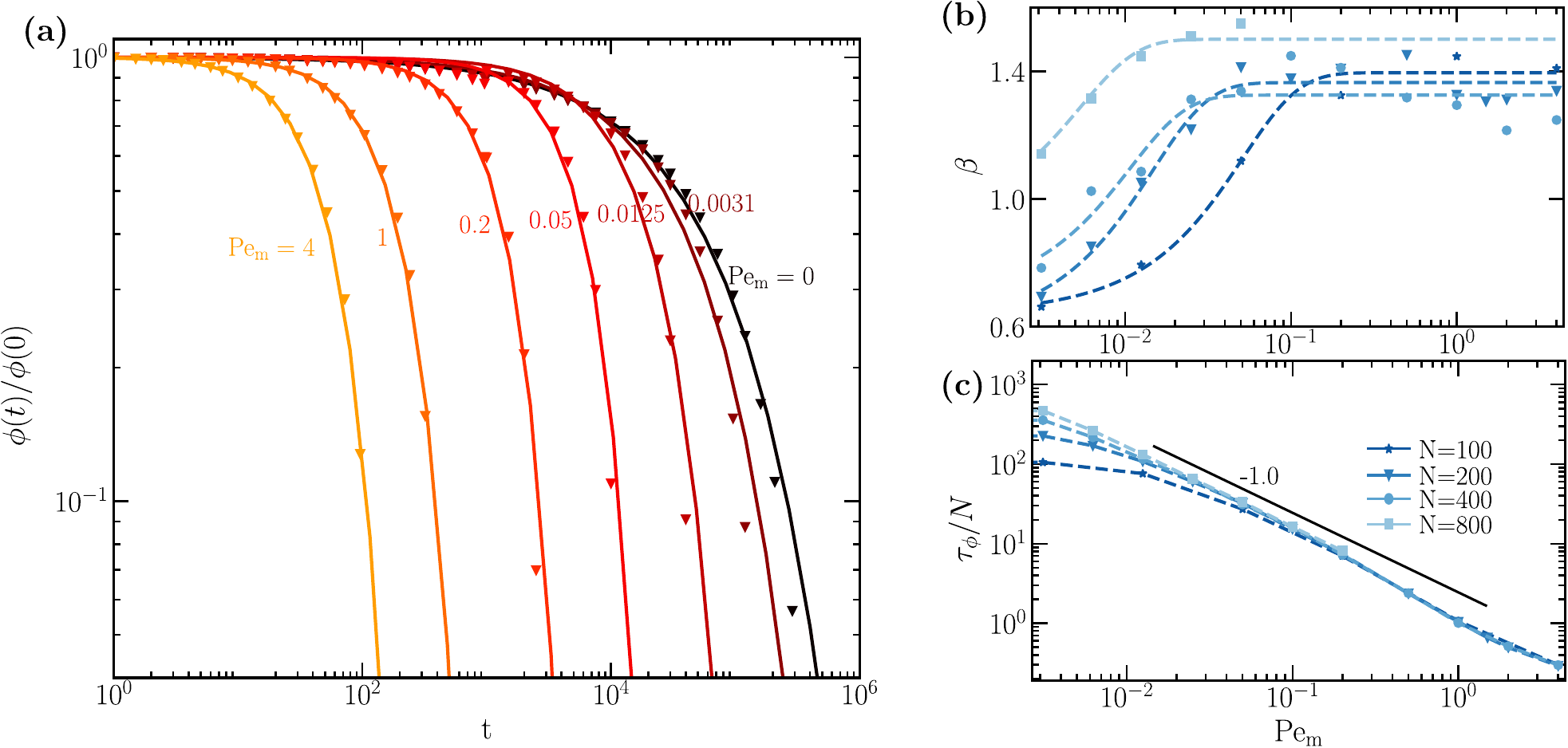}
	\caption{a)~Normalized auto-correlation function of the end-to-end vector for chains of $N=200$ and different activities. The lines are fits to Eq.~(\ref{eq:stretchexp}).
     b)~Stretching exponent $\beta$, and c)~characteristic relaxation time $\tau_\phi$, both as obtained from the fits in panel a), and calculated for different values of the activity and polymer length. 
      Dashed lines are guides to the eye and solid line in c)~depicts a power law of slope -1.} 
	\label{fig: Figphi}
\end{figure}

Relaxation of the end-to-end vector is one of the slowest dynamical processes in linear polymers,  and it can be experimentally measured by means of dielectric spectroscopy 
~\cite{Watanabe_1994, Watanabe_2005a}
Here, the end-to-end relaxation is measured through the auto-correlation function of the end-to-end vector,  
$\phi(t) = \langle \mathbf{R}(t_0) \mathbf{R}(t_0+t)\rangle$,
where the average is taken over all chains and time origins $t_0$, and results are
shown in Fig.~\ref{fig: Figphi}(a) for chains with molecular weight of $N=200$ and various activities. 
The passive case exhibits the slowest relaxation, following a nearly single-exponential behavior dominated by the slowest mode, as predicted by the tube theory~\cite{doi1988}. 
As polar activity increases, the chain drift along the tube accelerates, shortening the escape time, which explains that the end-to-end vector relaxes earlier the larger the activity. At higher activity values, the relative contribution of diffusion to reptation along the tube diminishes, causing a much sharper terminal region in the relaxation, as nearly all chains fully relax their orientation at the terminal time $\tau_c$ dictated by the drift.

To quantify the transition from diffusion-dominated to drift-dominated motion along the primitive path, we fit the end-to-end relaxation $\phi(t)$ to a stretched exponential, 
\begin{equation}
    \phi = e^{-\left(\frac{t}{\tau_\phi}\right)^\beta}
    \label{eq:stretchexp}
\end{equation}
where the fitting parameters $\beta$ (stretching exponent) and $\tau_\phi$ (relaxation time) are shown for different molecular weights and activities in Figs.~\ref{fig: Figphi}(b) and~(c), respectively. For the passive case, 
$\tau_\phi=\tau_d$, and when the activity dominates over reptation, $\tau_\phi=\tau_c$ (a comparison between the terminal times, as obtained from the center-of-mass MSD and the end-to-end relaxation, is shown in Fig. S3 of the Supplementary Material). 
For a single exponential relaxation process, $\beta$ should be equal to 1. 
In the passive case, $\beta\gtrsim 0.5$ since although the end-to-end relaxation is dominated by the first mode, higher odd modes also contribute~\cite{doi1988}. As the activity increases, $\beta$ grows to a value greater than 1, indicating a sharper relaxation process, as also shown in Fig.~\ref{fig: Figphi}(a). The relaxation time $\tau_\phi$ remains independent of activity for very small $\mathrm{Pe}_m$, when the diffusive reptation dominates over the minor drift caused by the activity. When activity overtakes reptation, the terminal time becomes inversely proportional to $\mathrm{Pe}_m$, consistent with expected scaling arguments~\cite{Tejedor_2020, tejedorProgressivePolymerDeformation2024}. 
Between Pe$_m=0.1$ and Pe$_m=1$, activity-induced bond alignment (see section \ref{subsec: OrderParameter}) reduces the effective monomeric friction coefficient, enhancing diffusion as shown in Fig. \ref{fig: FigDiffAndg3}(c), and slightly reducing the relaxation time of the end-to-end.

\subsubsection{Tube tangent correlation function and tube survival}

A central quantity in the development of tube theories is the tube tangent correlation function $G(i, i', t)$, which quantifies the correlation between the orientations of different primitive path segments located at positions $i$ and $i'$ along the tube, at different times. In Molecular Dynamics (MD) simulations, directly accessing the primitive path location is challenging, as it requires Primitive Path Analysis, a computationally intensive task if done frequently, and the \textit{tangent} vectors of the atomistic chain exhibit significant fluctuations. 
To address this, we define a coarse-grained version of the \textit{tangent-tangent} correlation function as:
\begin{equation}
G(i,i',t) = \langle \mathbf{t}(i,t) \cdot \mathbf{t}(i',0) \rangle,
\label{eq: tantan}
\end{equation}
where the coarse-grained \textit{tangent} vectors \(\mathbf{t}(i,t)\) are calculated by considering monomers spaced by 4 bonds (any number $n$ of bonds $1\ll n\ll N_e$ is expected to work). This method helps mitigate the fast decorrelation of the bond vectors due to CLF and lateral fluctuations within the tube:
\begin{equation}
\mathbf{t}(i,t) = \frac{\mathbf{r}_{i+4}(t) - \mathbf{r}_i(t)}{|\hat{\mathbf{t}}(i)|},
\end{equation}
where $i=0, 5, 10 \dots N$, $\hat{\mathbf{t}}(i)$ is the coarse grained tangent vector at equilibrium, and the average is taken over all the active chains in the system and across all possible time origins.

The primitive path can be viewed as a random walk, meaning that the tube-tangent correlation function at zero lag time corresponds to the equilibrium conformation of the tube and is expected to be nearly delta correlated, i.e. $G(i, i', t) \approx \delta(i-i')$. Due to the slithering motion along the tube, a segment $i$ at time $t$ may adopt an orientation similar to that of another segment $i'$ at time $0$. In pure reptation, the initially delta-correlated tangent-tangent function decays slowly by diffusion, while maintaining symmetry around the diagonals $i=i'$ and $i=N-i'$. Understanding the tube tangent correlation function enables the calculation of other critical observables, such as the stress tensor, structure factor, or tube survival function~\cite{grahamMicroscopicTheoryLinear2003}. In this work, we examine the impact of polar activity on the shape and decay of the tube tangent correlation function. 

\begin{figure}[hbt!]
	\centering
	\includegraphics[width=0.95\columnwidth]{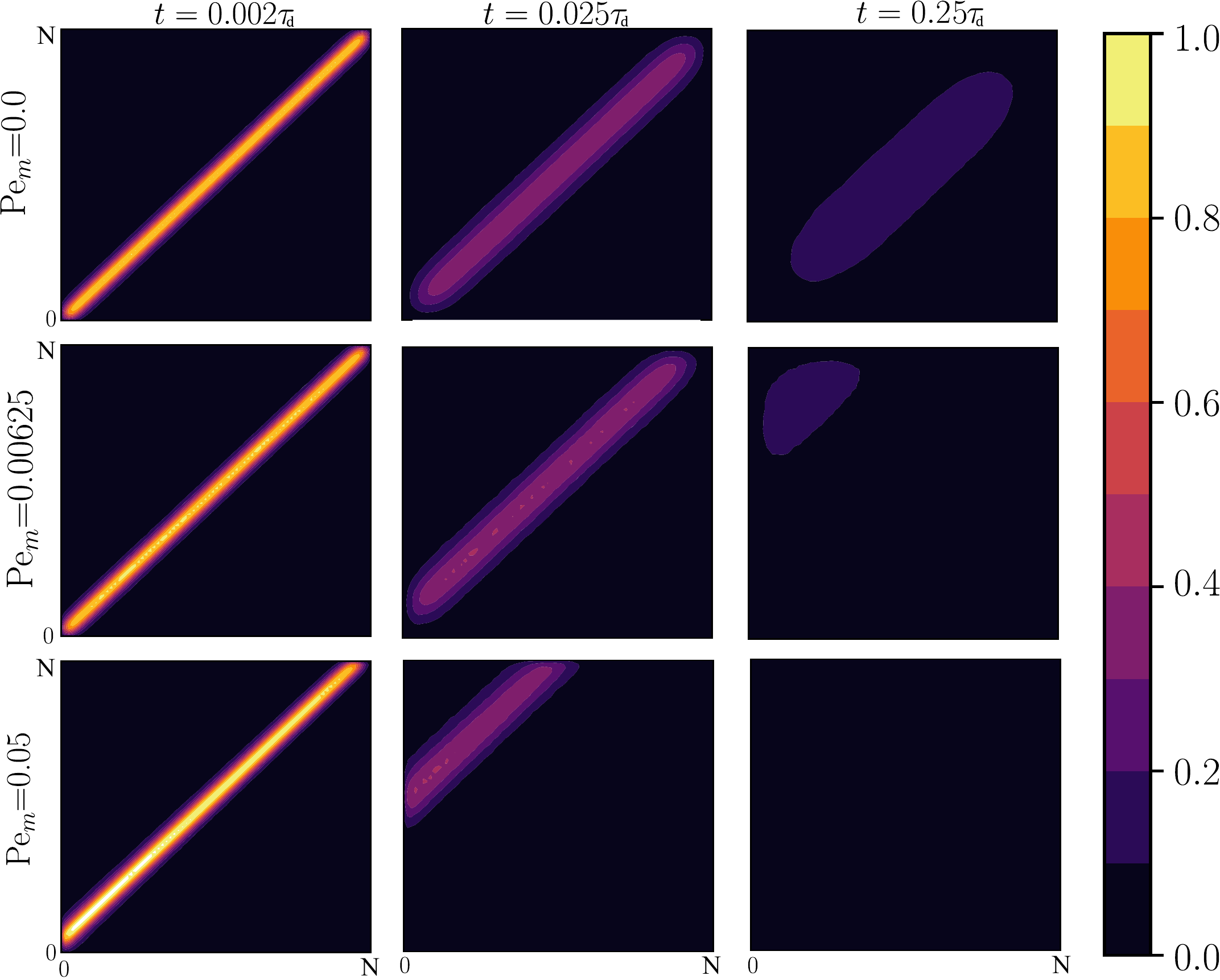}
	\caption{Tangent-tangent correlation function for chains of $N=400$ in the limit of low activities, evaluated at different times, with $\tau_d$ as obtained from Fig.~\ref{fig: Figphi}. The tail and head of the chain are represented by positions 0 and $N$, respectively. The passive case remains symmetric with respect to both diagonals at all times, decaying slowly by diffusion. For small activities the function remains symmetric only with respect to the diagonal $i=N-i'$ and shifts on time along that diagonal, indicating the drift of the tail segments towards the head. 
    }
	\label{fig: TanTan_slow}
\end{figure}
First, we focus on a range of small activities, Pe$_m\lesssim 0.05$, where the theory of active reptation holds. In Fig.~\ref{fig: TanTan_slow} the tangent-tangent correlation function is shown for $N=400$ at three different activities and various lag-times. At very early times, the function is nearly zero everywhere except along the diagonal. In the passive case (top row in Fig.~\ref{fig: TanTan_slow}), the function remains symmetric and decays slowly, with maximum values consistently aligned with the diagonal, as predicted by the tube theory. As $t$ approaches the terminal time $\tau_d$, the function decays to zero. For small activities (bottom rows in Fig.~\ref{fig: TanTan_slow}), diffusion-like reptation motion dominates at early times, and the tangent-tangent function remains symmetric. However, as time progresses, the function shifts diagonally toward the upper-left corner, and the maximum values are no longer located along the diagonal, breaking the head-tail symmetry. Polar activity induces a systematic drift along the tube, where the orientation of tail segments (position 0 on both axes) at time $t$ correlates with head segments (position $N$ on both axes) from an earlier time. This occurs because the tail segments at time $t$ pass through tube segments previously occupied by the head at time 0. As activity increases, this diagonal shift becomes more pronounced, and the decay of the tangent-tangent function accelerates, indicating that the chain is escaping the tube much faster than by reptation alone.

\begin{figure}[hbt!]
	\centering
	\includegraphics[width=0.95\columnwidth]{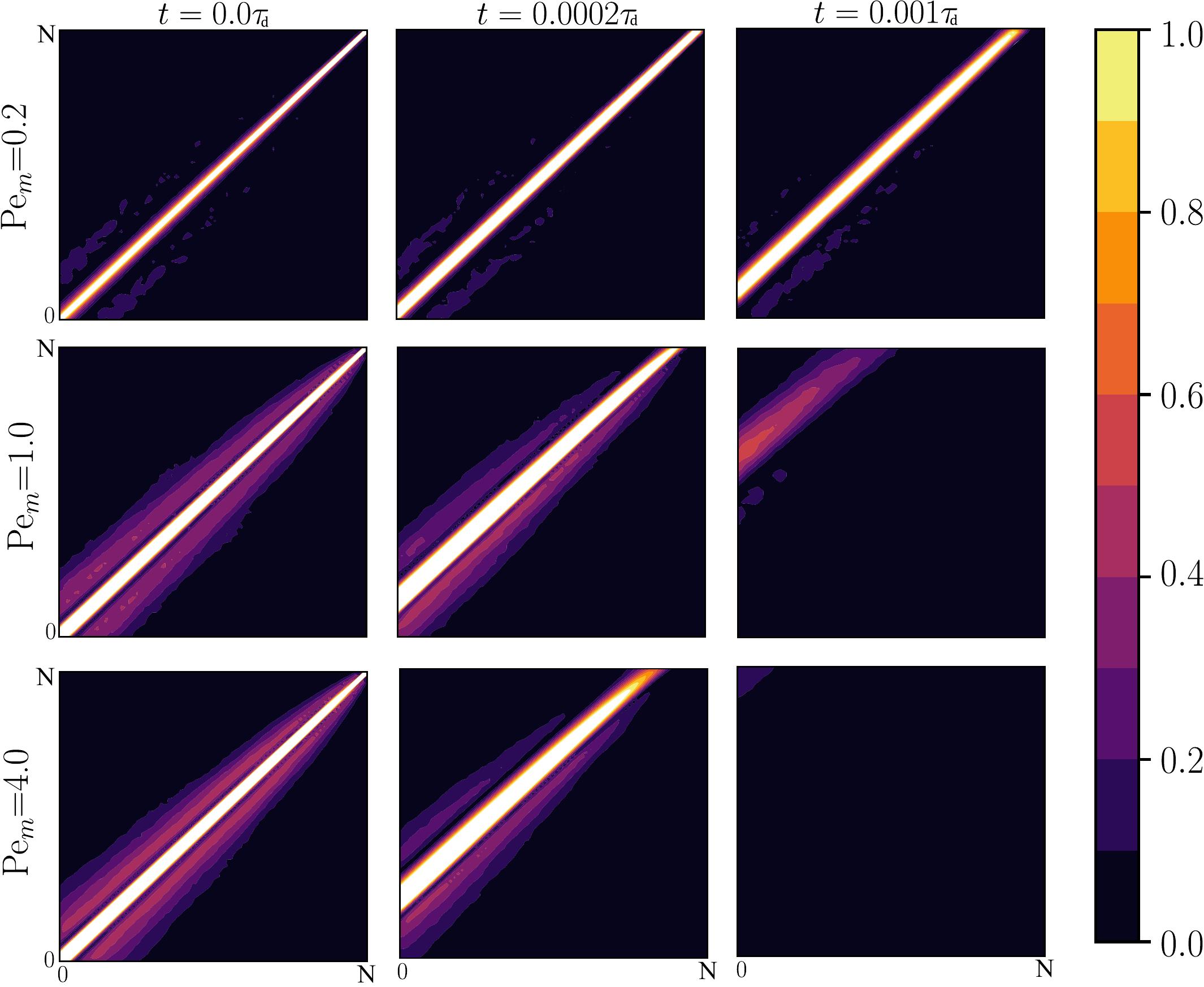}
	\caption{Tangent-tangent correlation function for chains of $N=400$ in the limit of high activities, evaluated at different fractions of the disengagement time $\tau_d$ (see previous Figure). Head-tail symmetry is broken, so the function is not symmetric with respect to any diagonal. The drift of the tail segments towards the head occurs at much shorter times. }
	\label{fig: TanTan_fast}
\end{figure}
As discussed in previous sections, when the activity exceeds the range of validity of the active reptation theory, additional effects such as segmental and tube elongation, orientational correlation of tube segments, or bond alignment emerge. This is reflected in the shape of the tube-tangent function, as shown in Fig.~\ref{fig: TanTan_fast}. The tube conformation is no longer at equilibrium, and the tube-tangent function is not delta-correlated at time 0. $G(i,i',0)$ exhibits a halo that intensifies as the segment approaches the tail, indicating stronger correlation between segments due to activity-induced tail alignment. Additionally, the values along the diagonal exceed 1, signaling segment elongation with respect to equilibrium. In contrast, head segments remain nearly delta-correlated. As time progresses, the tangent-tangent function decays rapidly as the chain relaxes and loses memory of its previous orientation. 

\begin{figure}[hbt!]
	\centering
	\includegraphics[width=\columnwidth]{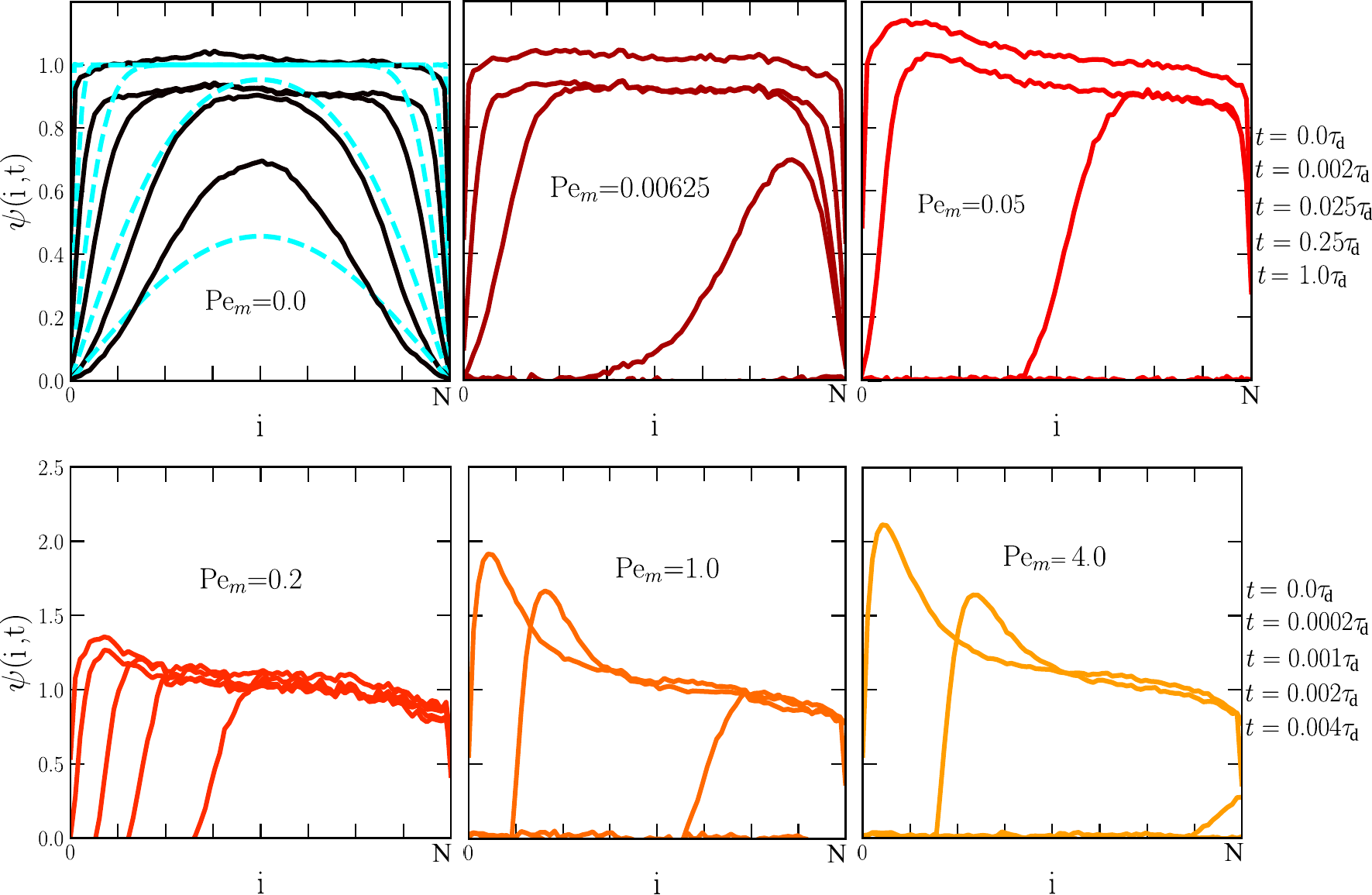}
	\caption{Tube segment survival function for chains with $N=400$ at different Pe$_m$ values,
    where $i$ represents the position of the tangent vector along the polymer. The times specified at the right indicate the times at which each of the lines has been evaluated. 
    The cyan dashed lines in the first panel show the prediction of the standard reptation theory (Eq. 6.14 in the book of Doi and Edwards~\cite{doi1988}) for the tube segment survival function.}
	\label{fig: Surfun}
\end{figure}
The tube segment survival function $\psi(i, t)$ is another key component of the tube theory, representing the probability that a segment $i$ along the tube has not yet been visited by either end of the chain at time $t$. The function $\psi(i,t)$ is calculated by integrating the tube-tangent correlation function $G(i, i', t)$ with respect to either index $i$ or $i'$, due to the symmetry of the function along the secondary diagonal. 
In equilibrium, the tube theory predicts that $\psi(i,t)$ exhibits head-tail symmetry and decays slowly at both ends, since both the head and the tail can destroy the tube through reptation. According to the theory of active reptation~\cite{Tejedor2019,Tejedor_2020}, the activity-induced drift introduces an asymmetry in $\psi(i,t)$, with tube segments being renewed more quickly by the tail than by the head, as well as a much faster decay of the tube survival function. These predictions have been confirmed through single-chain Brownian Dynamics simulations~\cite{Tejedor_2020} and multichain MD simulations of active chains in a mesh of passive long chains~\cite{tejedorMolecularDynamicsSimulations2023}, considering CLF. Here, we test whether the predictions hold when the effect of CR is significant. As shown in Fig.~\ref{fig: Surfun}, at low activity levels, the MD simulation results align with theoretical predictions. 
As the activity increases, the effects of reptation and CLF diminish, and the decay of $\psi(i,t)$ is primarily driven by the polar drift motion of the chain along the tube and towards the head. At low values of the activity this results in a sharp decay of the tube survival function at the tail and, as shown in Fig.~\ref{fig: Surfun}, $\psi(i,t)$ shifts to the right as time proceeds. At high Pe$_m$, due to the stretching of tail segments, $\psi(i,t)$ exceeds the value 1 at positions near the tail, and these maxima are dampened and shifted to towards the head as chain drifts along the tube.



\subsection{Effect of Inertia}\label{subsec: Inertia}

In polymer melts at equilibrium, inertia effects are generally negligible, but the introduction of activity may alter this. To better understand the influence of inertia, we conduct additional simulations using Langevin dynamics as in Eq.~(\ref{eq:lang}), using a higher friction value to reduce the eventual relevance of inertia.

We first estimate the effective equilibrium monomer friction $\zeta_0$, which may differ from the input friction coefficient $\zeta$ specified in Eq.~(\ref{eq:lang}). While $\zeta$ and $\zeta_0$ should be equivalent in dilute systems, collisions between nearby monomers in dense systems can alter the effective friction. Long simulations for $N=200$ chains are conducted, and the diffusion coefficient $D_{G0}$ is measured via the center of mass MSD; values are shown in Table~\ref{tab:DG0}. Although chains of this length have only 3-4 entanglements, we estimate the monomeric friction coefficient using the Rouse expression $D_{G0}=k_BT/N\zeta_0$. Acknowledging that this approach may slightly overestimate friction due to the presence of entanglements, the results in Table~\ref{tab:DG0} suggest that the effective monomer friction $\zeta_0$ increases roughly five times less than the input friction $\zeta$, likely due to increased monomer collisions at lower $\zeta$ values. 
\begin{table}[]
    \centering
    \begin{tabular}{|c|c|c|}
         \hline
         $\zeta$ &  $D_{G0}$ & $\zeta_0$\\
         \hline
         0.5&  5.5e-4 & 9.1 \\
         20 & 7.2e-5 & 69.4 \\
         \hline
    \end{tabular}
    \caption{Diffusion coefficient at equilibrium for different values of the input friction parameter $\zeta$ used in Eq.~(\ref{eq:lang}), and the related effective monomeric friction coefficient $\zeta_0$, which has been estimated from Rouse theory by $D_{G0}=k_BT/N\zeta_0$.}
    \label{tab:DG0}
\end{table}

\begin{figure}[hbt!]
    \centering
    \includegraphics[width=1.0\columnwidth]{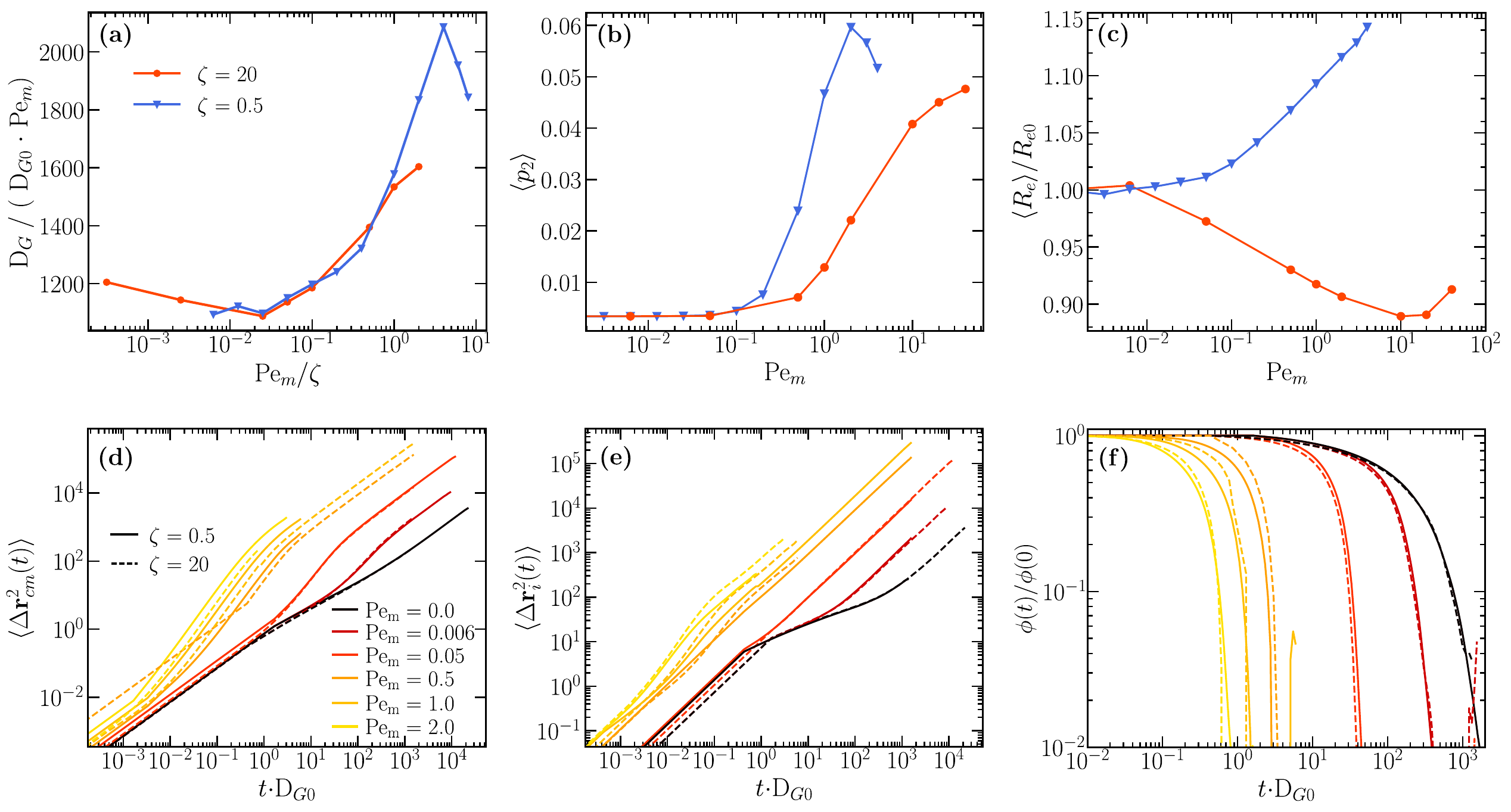}
	\caption{Effect of inertia for chains of $N=200$: (a)~Diffusion coefficient, (b)~bond alignment~$p_2$ , (c)~normalized end-to-end size, (d)~center of mass mean square displacement, (e)~monomeric mean square displacement, and (f)~end-to-end relaxation. \\
    }
	\label{fig: Figinertia}
\end{figure}

When polar activity dominates, according to theoretical predictions~\cite{Tejedor2019}, the diffusion coefficient should be proportional to the drift velocity along the primitive path, $c$, which is inversely proportional to the monomeric friction coefficient $\zeta_0$. Therefore, in the presence of activity, the ratio of the diffusion coefficients should be inversely proportional to the ratio of the monomeric frictions. 
Fig. \ref{fig: Figinertia}(a) shows the diffusion coefficient $D_G$, normalized by its equilibrium value $D_{G0}$ and the activity, plotted against Pe$_m$ scaled by 
$\zeta$. Both curves agree at low activities, supporting the proportionality between $D_G$ and Pe$_m$~
\cite{Tejedor2019}. 
At high activities, both curves exhibit a similar upturn, likely due to activity-induced bond alignment (see Sec.~\ref{subsec: OrderParameter}). Fig.~\ref{fig: Figinertia}(b) compares bond alignment for both frictions,  showing a qualitatively similar increase in both cases, although alignment occurs at a higher activities as $\zeta$ increases. 

The evaluation of coil size as a function of of Pe$_m$ is shown in Fig. \ref{fig: Figinertia}(c) for the two investigated values of $\zeta$, revealing notable qualitative differences.  
For higher friction, coil size decreases with Pe$_m$, consistent with recent results obtained for the same system~\cite{ubertiniUniversalTimeLength2024}. In contrast, for lower friction, chains expand monotonically with activity, as discussed in Sec.~\ref{subsec:RG}. 
Similar to the behavior observed in dilute chains~\cite{tejedorProgressivePolymerDeformation2024}, coil size  shows a clear upturn at activities beyond Pe$_m/\zeta >1$, where chain size trends change.
In simulations with smaller friction, coil size does not initially decrease, but instead increases at activities above Pe$_m \approx 0.1$

To evaluate the impact of friction on system dynamics, Figs. \ref{fig: Figinertia}(d), (e) and (f) show the mean-square displacements of the center-of-mass and middle monomer, as well as the end-to-end vector relaxation, respectively, 
all plotted against time normalized by $b^2/D_{G0}$, the time required for the chain to move a distance on the order of the bead size at equilibrium. Dynamic properties are nearly independent of $\zeta$, showing that both systems exhibit similar behavior,
with differences emerging primarily from time shifts based on $\zeta$. Discrepancies appear only at high Pe$_m$, where bond alignment becomes more pronounced at lower friction, reducing the effective monomeric friction and thereby enhancing relaxation and diffusion. Bond deformation due to activity, as shown in Fig. S5 of the Supplementary Material is also notable: bonds deform significantly at Pe$_m=1$ for $\zeta=0.5$ and Pe$_m=10$ for $\zeta=20$. Thus, at high Pe$_m$, our results should be interpreted with caution, as the uncrossability of the chains might be compromised. However, the extent of bond stretching (10\%) remains limited enough to preserve the validity of our results at high Pe$_m$. 

\subsection{Phase Diagram}\label{subsec: PhaseDiag}

The complex dynamical and conformational behavior of active polar polymer chains in the melt can be summarized in the phase diagram shown schematically in Fig.~\ref{fig: FigPhaseDiagramSketch}(a), and particularized for the Kremer-Grest model in Fig.~\ref{fig: FigPhaseDiagramSketch}(b). The diagram is expressed as a function of molecular weight $N$ and P\'{e}clet number Pe$_m$. In this diagram, eight distinct regions can be identified, considering the relative effects of molecular weight, entanglement, activity, anisotropy, chain and tube stretching, and bond alignment. The transitions between adjacent regions are not sharp and may occur over a broad crossover; however, points well within each region should exhibit representative behavior characteristic of that region. Using the scaling behaviors previously discussed, the positions of each boundary line separating the regions are determined as follows:
\begin{figure}[hbt!]
	\centering
    \includegraphics[width=\columnwidth]{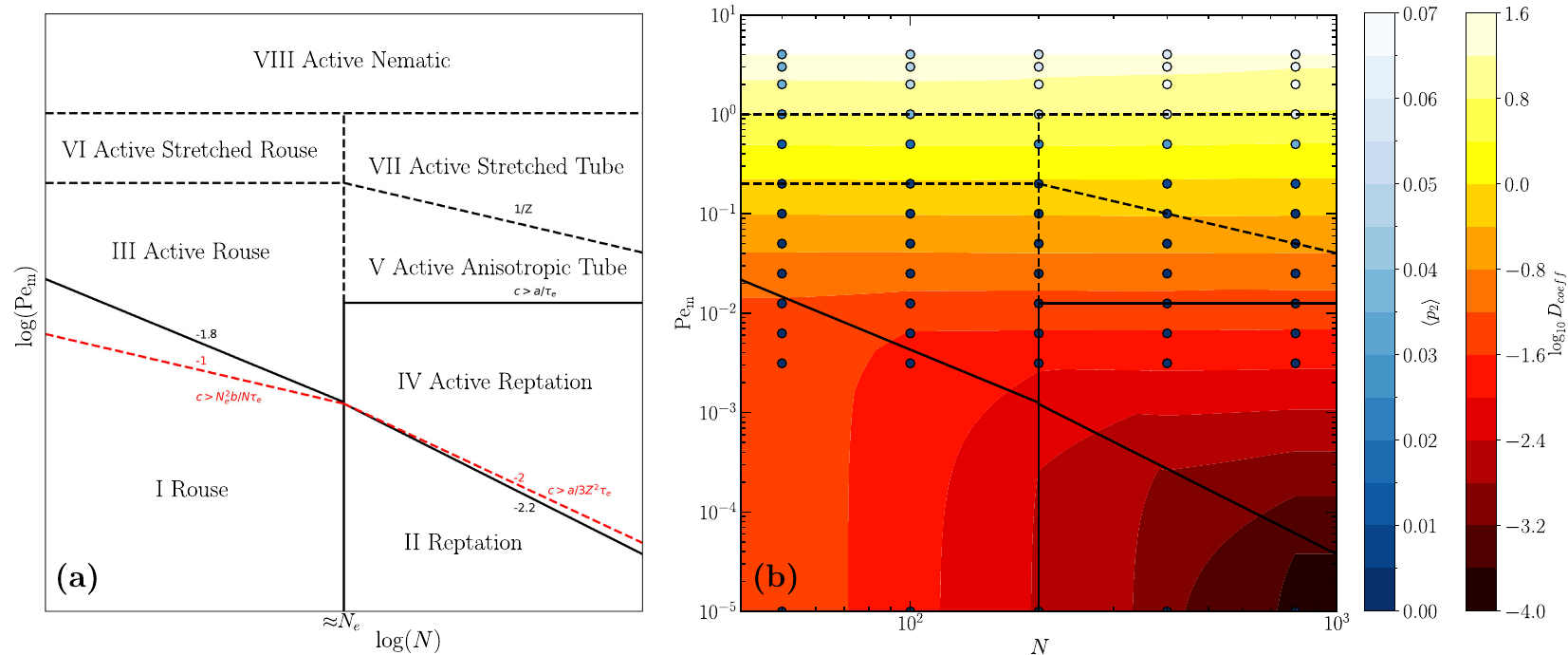}
	\caption{Phase diagram of active polar linear polymer melts: (a) Diagram showcasing the different dynamical and conformational regions; (b) Diagram with simulation results with color field representing the diffusion coefficient, and the color of the markers based on the average nematic order of the system. }
	\label{fig: FigPhaseDiagramSketch}
\end{figure}

\begin{itemize}
    \item The vertical line at $N\approx N_e$ delineates the boundary between unentangled (region I) and entangled (region II) chains. This transition is typically observed at 2-4 times $N_e$. This line is continuous up to $c\le a/\tau_e$, where $c$ is the drift velocity along the primitive path imparted by the tangent polar activity (which is proportional to Pe$_m$) \cite{Tejedor2019}. Beyond this line, the assumptions of the tube theory no longer apply, and at even higher activities, it becomes uncertain whether surrounding chains continue to constrain the probe chain.  
    \item The line separating reptation (II) from active reptation (IV) follows from theoretical predictions~\cite{Tejedor2019}, which indicate a critical activity that scales as $c>a/3Z^2\tau_e$, where $Z$ is the number of entanglements. Above this threshold, the activity-induced drift dominates over the diffusive reptation motion. In our simulation results, we identify the threshold when the activity-enhanced diffusion coefficient is 1.5 greater than the equilibrium value due to pure reptation and CLF (see Fig.~\ref{fig: FigDiffAndg3}). The resulting slope is -2.2, rather than the predicted -2, likely due to the influence of CLF on diffusion.
    \item The limit of validity of the active reptation theory separates regions IV and V and is reached when $c\approx a/\tau_e$, a condition that is independent of molecular weight. When the drift velocity surpasses $a/\tau_e$, new tube segments are created by the head without permitting the exploration of all possible orientations. Beyond this threshold, the isotropy of newly created tube segments can no longer be assured, leading to correlated orientation among tube segments (see Fig.~\ref{fig: Figbondcorr2}).
    \item The line separating the Rouse regime(I) from the Active Rouse regime (III) can be determined by comparing the time it takes for the tail monomer to drift over the whole chain contour ($\tau_c = Nb/c$) with the Rouse time for a chain of molecular weight $N$, $\tau_R = N^2\tau_e/N_e^2$. When $\tau_c < \tau_R$, the activity dominates over Rouse relaxation, which translates to $c>N_e^2b/N\tau_e$. This line intersects both the boundary between unentangled (I) and entangled (II) polymers and the boundary separating pure reptation (II) from active reptation (IV) at a single point. The slope of this line is -1.8 rather than -1, attributed to the dependence of the diffusion coefficient on molecular weight in simulations of unentangled polymers (see Section \ref{subsec: MSD of COM}). 
    \item As discussed in section \ref{subsec: PPA}, high activity levels may cause tube stretching. The line separating active anisotropic reptation (V) from the active stretched tube (VII) can be derived from Fig.~\ref{fig: FigPPA} by establishing a threshold elongation of the tube at 10\%. The global P\'{e}clet number corresponding to this relative elongation of the tube is Pe$_g\approx 100$. Consequently, this leads to the relationship Pe$_m\propto 100/N \propto 1/Z$. 
    \item Similarly, there is a critical activity above which unentangled chains also experience elongation (region IV). The limiting P\'{e}clet number corresponding to this transition can be extracted from Fig.~\ref{fig: FigRe} as Pe$_m\approx0.2$. This transition line is likely dependent on molecular weight and should intersect the line separating unentangled (I) from entangled (III) chains, as well as the line that separating anisotropic (V) from stretched (VII) tube at a single point. 
    \item The line separating regions VI and VII from region VIII is independent of molecular weight and can be extracted from Fig.~\ref{fig: Figp2}. In region VIII, the activity induces nematic bond alignment. 
\end{itemize}


In Fig.~\ref{fig: FigPhaseDiagramSketch}(b), the diffusion coefficient (depicted as a background color field) and the mean value of the nematic order parameter $\langle p_2\rangle$ (represented by the intensity of the colored symbols, ranging from black to white) are presented to support the description in terms of a phase diagram.  For instance, the color bands are vertical in regions I and II, indicating that the diffusion coefficient depends solely on the molecular weight of the polymer. However, this bands become horizontal in the rest of the phase diagram, reflecting the dependence of diffusion on activity. The nematic order is nearly negligible below the line Pe$_m$=1, and it becomes maximum at the top of the diagram. Other quantities examined in this work, such as chain stretch and orientation correlation, exhibit distinct behaviors accross different regions of the phase diagram.

\section{\label{sec:conclusion}Summary and Conclusions}

In this work, we investigated the effect of tangent polar activity on the conformational and dynamical behavior of linear polymer melts through Langevin dynamics, using the Kremer-Grest model. The fact that all chains in the system are active allows us to explore the effects of constraint release (CR), which were not explored in our previous work \cite{tejedorMolecularDynamicsSimulations2023}. 
The main findings of our study can be summarized as follows: 
\begin{itemize}
    \item The overall polymer conformation, as represented by the end-to-end vector, is slightly stretched due to the activity, which contrasts with the trend observed for the same active model in dilute conditions \cite{tejedorProgressivePolymerDeformation2024}. 
    Simulations using a higher friction coefficient also show a more compact structure, in accordance to recent results~\cite{ubertiniUniversalTimeLength2024}.
    This difference can therefore be attributed to inertia effects, which become significant at high activity levels. 
    \item The change in conformation shows a universal behavior across different molecular weights when plotted as a function of the global P\'{e}clet number, up to a high threshold activity, where deviations begin to occur.
    \item Similar to dilute chains, the deformation is not homogeneous along the chain contour, with heads being more compact and tails more elongated, though different molecular weights do not follow a universal trend. 
    \item The entanglement network remains unperturbed up to a threshold Pe$_m=0.0125$, beyond which, orientational correlation between entanglements appear, and the primitive path becomes elongated. 
    \item Above Pe$_m=0.1$, activity-induced bond alignment is observed, growing to a maximum at Pe$_m=2$ and decays at larger activity levels. Large dynamic clusters of aligned bonds emerge within the system, but rapid fluctuations prevent phase separation.
    \item The monomeric and center-of-mass diffusion can be well described by the theory of active reptation up to a limiting activity of Pe$_m=0.0125$, beyond which the assumptions of reptation no longer apply. 
    \item In the range of validity of the theory, the center of mass MSD shows superdiffusive behavior, with the diffusion coefficient becoming independent of molecular weight and scaling linearly with Pe$_m$. At higher activity, the diffusion coefficient grows more rapidly due to bond alignment, which reduces friction. As in dilute chains, the center of mass MSD can be described accurately by the equation proposed for the MSD of ABPs.  
    \item Polar activity breaks the dynamical symmetry between the head and tail, with the head monomer becoming the slowest, while the tail becomes the fastest monomer of the chain. At high activities, the monomeric MSD also shows a superdiffusive regime. 
    \item As activity increases, the end-to-end relaxation occurs more rapidly, and the decay becomes shaper, with the relaxation time scaling as $N/$Pe$_m$. 
    \item The tube tangent correlation function and tube survival function align with the predictions of the active reptation theory up to Pe$_m=0.0125$. At higher activities, the stretching and orientation correlation of the tube segments cause the simulation results to diverge from the theoretical predictions. 
    \item The overall complex dynamical and conformational behaviour can be summarized in a phase diagram as a function of Pe$_m$ and $N$. 
\end{itemize}

Our work contributes to the development of a robust theoretical framework to investigate the dynamics of active polar polymers, which will be invaluable for advancing both experimental research and theoretical models in the field. The findings of this study can serve as a foundation for exploring active biomolecules and guiding the design of new macromolecular materials with enriched and enhanced dynamical properties.

\section*{Acknowledgements}

The authors acknowledge funding from the Spanish Ministry of Economy
and Competitivity (PID2019-105898GA-C22 and PID2022-136919NB-C32) and the Madrid
Government (Comunidad de Madrid-Spain) under the Multiannual
Agreement with Universidad Politécnica de Madrid in the line Excellence
Programme for University Professors, in the context of the V PRICIT
(Regional Programme of Research and Technological Innovation). A.R.T. acknowledges funding from EMBO scientific exchange grants and EBSA bursary. 
The authors gratefully acknowledge the Universidad Politécnica de Madrid (\url{www.upm.es}) for providing computing resources on Magerit Supercomputer.
The authors gratefully acknowledge the computing time granted by the JARA Vergabegremium and provided on the JARA Partition part of the supercomputer JURECA at Forschungszentrum J\"ulich~\cite{jureca2018}.









\end{document}